\documentclass[twocolumn]{article}
\usepackage[left=2.5cm, right=2.5cm, top=2.5cm, bottom=2.5cm]{geometry}
\usepackage[utf8]{inputenc}
\usepackage[british]{babel}
\usepackage{authblk}
\usepackage{amsmath, amssymb}
\usepackage{graphicx}
\usepackage{xcolor}
\usepackage[colorinlistoftodos]{todonotes}
\usepackage{textcomp}
\usepackage{hyperref}
\usepackage[ruled,vlined]{algorithm2e}
\usepackage{bbm}

\newcounter{comment}

\newcommand{\ABcolor}{violet} 
\newcommand{\JBcolor}{blue}
\newcommand{\HMcolor}{red}

{\refstepcounter{comment}%
\begin{quote}
\color{\ABcolor}\small$\blacksquare$ \textbf{\underline{Comment} $\sharp$\thecomment.~AB:}}%
{\end{quote}}

{\refstepcounter{comment}%
\begin{quote}
\color{\JBcolor}\small$\blacksquare$ \textbf{\underline{Comment} $\sharp$\thecomment.~JB:}}%
{\end{quote}}

{\refstepcounter{comment}%
\begin{quote}
\color{\HMcolor}\small$\blacksquare$ \textbf{\underline{Comment} $\sharp$\thecomment.~HM:}}%
{\end{quote}}

\DeclareMathOperator*{\argmin}{Arg\!min}
\DeclareMathOperator{\Tr}{Tr}
\newcommand{\norm}[1]{\left\lVert#1\right\rVert}
\newcommand{\normBlock}[1]{{\left\vert\kern-0.25ex\left\vert\kern-0.25ex\left\vert #1 \right\vert\kern-0.25ex\right\vert\kern-0.25ex\right\vert}}

\providecommand{\keywords}[1]{\textbf{\textit{Keywords---}} #1}

\newcommand{\nonparametric}{non-parametric\xspace}
\newcommand{\quasimonochromatic}{quasi-monochromatic\xspace}
\newcommand{\quasiperiodic}{quasi-periodic\xspace}
\newcommand{\autocorrelation}{auto-correlation\xspace}
\newcommand{\nonzero}{non-zero\xspace}
\newcommand{\BlockTree}{\texttt{BlockTree}\xspace}
\newcommand{\eg}{\textit{e.g.}\xspace}
\newcommand{\ie}{\textit{i.e.}\xspace}

\title{Sparsity Based Recovery \\ of Galactic Binaries Gravitational Waves}
\author[1]{A.~Blelly\footnote{aurore.blelly@cea.fr}}
\author[1,2]{J.~Bobin}
\author[1]{H.~Moutarde}
\affil[1]{IRFU, CEA, Université Paris-Saclay, F-91191 Gif-sur-Yvette, France}
\affil[2]{IPARCOS, Complutense University, E-28040 Madrid, Spain}

\date{\today}

\begin{document}

\maketitle

\begin{abstract}
The detection of galactic binaries as sources of gravitational waves promises an unprecedented wealth of information about these systems, but also raises several challenges in signal processing. In particular the large number of expected sources and the risk of misidentification call for the development of robust methods. We describe here an original \nonparametric reconstruction of the imprint of galactic binaries in measurements affected by instrumental noise typical of the space-based gravitational wave observatory LISA. We assess the impact of various approaches based on sparse signal modelling and focus on adaptive structured block sparsity. We carefully show that a sparse representation gives a reliable access to the physical content of the interferometric measurement. In particular we check the successful fast extraction of the gravitational wave signal on a simple yet realistic example involving verification galactic binaries recently proposed in LISA data challenges.
\end{abstract}

\keywords{Galactic binaries, gravitational waves, \quasiperiodic signals, sparse signal representation, \nonparametric signal estimation, LISA mission, reweighted L1 minimization, block decomposition, LISA Data Challenges}


\section{Galactic binaries detection and gravitational waves}

\subsection{Introduction}

Gravitational waves (GWs) are distortions of the space-time curvature propagating at light speed. They can be emitted by a great variety of sources: black holes mergers, supernovae, extreme-mass ratio inspirals, galactic binaries (GBs), \textit{etc}. They can now be detected using interferometers such as the devices from the LIGO-Virgo collaboration, and later will be observed directly from space thanks to the LISA (Laser Interferometer Space Antenna) project \cite{Audley:2017drz} led by the European Space Agency.

LISA is a giant interferometer made up of 3 satellites forming 3 arms 2.5 millions kilometers long. These 3 arms produce time series on 3 main data streams, called TDI X, Y, Z, where TDI stands for time delay interferometry. By linearly combining these data streams, we obtain the 3 channels A, E and T with maximum signal-to-noise (SNR) ratio and theoretically uncorrelated noises (see the review \cite{Tinto:2014lxa} and references therein).

In order to identify the gravitational signals acquired by these data channels, many methods have already been proposed, especially in the frame of the LIGO-Virgo collaboration. However, they cannot be straightforwardly implemented for the LISA mission. Indeed the expected measurements are fundamentally different than those made in LIGO-Virgo, from the amount of sources that the device could potentially detect to their frequency range (not even mentioning the unique imprint of glitches, instrumental noise or gaps). In particular, LISA will permanently monitor millions of GBs --white dwarfs, neutron stars or stellar-origin black holes-- emitting \quasimonochromatic GWs. These numerous sources will induce a stochastic foreground that will impact most of the channels investigated by LISA. The science objectives of the mission specify the resolution and characterization of several thousands of GBs. LISA will deliver an unprecedented wealth of information about GBs. For example, 16 ultracompact binary systems, coined \emph{verification binaries} (VBs), have been electromagnetically identified \cite{Kupfer:2018}. Thanks to Gaia \cite{GaiaColl:2016} and LSST \cite{LSSTScienceColl:2009}, a known population of about 100 to 1000 VBs is foreseen \cite{Korol:2017qcx} before LISA starts taking data. The expected detection of tens of thousands of white dwarf systems during LISA lifetime (see Ref.~\cite{Lamberts:2019nyk} and references therein) will pave the way for key statistical analysis. Making the full benefit of it call for the development of new analysis techniques to address the associated data processing challenges.

We hereby propose a \nonparametric\footnote{We could equivalently call it a \emph{model-independent} method: we will not evaluate the physically meaningful parameters of a specific model but we will rather approximate a measured signal by an expansion on an suitable basis.} method for GB detection through sparse recovery. Sparse signal processing has emerged as a major methodology in signal processing during the last two decades. More precisely, it sparkled the development of new analysis methods to solve challenging ill-posed or ill-conditioned inverse problems \cite{Bruckstein2009}. Sparse signal modelling is at the origin of some of the state-of-the-art methods for standard inverse problems ranging from denoising to compressed sensing to only name two \cite{starck:book10}. It has already been applied to the detection by LIGO-Virgo of gravitational waves emitted by mergers of black holes as \eg in Refs.~\cite{Bacon:2018zgb, Feng:2018}.

In our case, we apply sparse signal modelling to GBs from a different perspective. As the GB gravitational waveforms are \quasimonochromatic, their frequency representation can be thought of as begin sparse in the classical Fourier basis as shown in Fig.~\ref{fig:GB_waveform}. We therefore propose to recast GB detection and reconstruction as a particular denoising problem, where sparse signal modelling allows for increased robustness with respect to noise.

In this article we establish a new methodology, based on denoising and sparse signal recovery, which enables the detection and the reconstruction of GW signals coming from GBs.  The basic framework is detailed in Section~\ref{par:FbF_sparsity}. Based on an innovative tree-based block signal decomposition in the harmonic domain, it allows to precisely control the statistics of the detection process, namely the false positives and false negatives rates. The proposed structured sparsity framework is detailed in Section~\ref{par:block_sparsity}. In Section~\ref{par:performances_benchmark} all these algorithms are benchmarked on data created with a fast waveform generator provided by the LISA Data Challenge team. Section. Conclusion will be drawn in Section~\ref{par:Conclusion}.

\subsection{Problem statement}
\label{par:context}

\begin{figure*}[h!]
    \centering
    \includegraphics[width=1\textwidth]{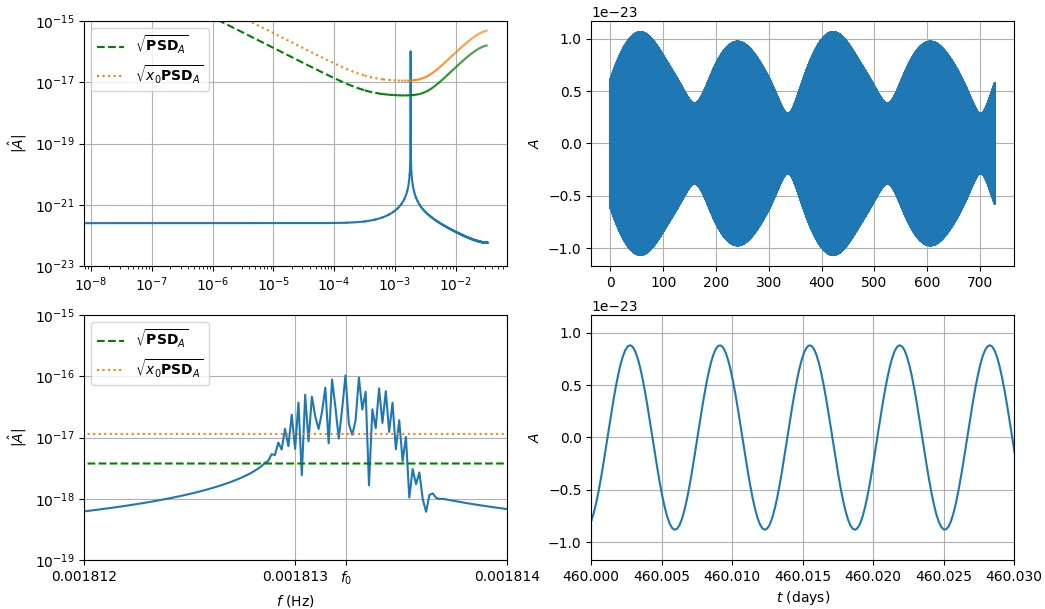}
    \caption{GB waveform with central frequency $f_0 = 1.81324$ mHz. Left column: Frequency domain; Right column: time domain. Top: full signal; Bottom: zooms. On the left column, $\sqrt{\mathbf{PSD}_A}$ (dashed green) represents the expected noise PSD and $\sqrt{x_0 \mathbf{PSD}_A}$ (dotted orange) represents the threshold for which $99\%$ of noise realisations have a lower amplitude. In time domain, we observe that the high frequency signal, which is the one that we seek, admits a very low frequency envelop.}
    \label{fig:GB_waveform}
\end{figure*}

As exemplified by Fig.~\ref{fig:GB_waveform} in frequency domain, the waveform is essentially monochromatic. The zoom displays a wide frequency peak where the signal can oscillate up to 1 order of magnitude. Still only few frequencies get past the noise level. The higher the chosen threshold, the fewer the frequencies will be detected as "active" (in a sense which will be made precise in Section~\ref{par:L1_minimization}). This justifies the relevance of a sparse modelling of the waveform in Fourier domain.

However such a continuous waveform is not exactly what will be experimentally retrieved. One GW reaching the satellite constellation will instead generate a signal on each of the 3 Michelson-Morley-like arms of the detector. LISA will thus deliver 3 time series: the 3 TDI signals $X$, $Y$,$Z$, sampled at time $t_n = n \Delta T$ ($0 \leq n < N_T$). As mentioned above, by linearly combining them, we get the 3 data channels $A$, $E$ and $T$: 
\begin{eqnarray}
A[n] &=& \frac{Z[n] - X[n]}{\sqrt{2}} \;, \label{eq:def-channel-A} \\
E[n] &=& \frac{Z[n] - 2 Y[n] + X[n]}{\sqrt{6}} \;, \label{eq:def-channel-E} \\
T[n] &=& \frac{Z[n] + Y[n] + X[n]}{\sqrt{3}} \;, \label{eq:def-channel-T} 
\end{eqnarray}
whose noises are assumed to be statistically uncorrelated. Since the SNR of channel T is much smaller than those of the channels A and E in the mHz domain \cite{Prince:2002hp} --which is the relevant range for GB detections by LISA-- we will neglect $T$ and consider only the signals from A and E \cite{Robson:2018jly}. 

For the sake of conciseness, we will note: 
\begin{equation}
\label{eq:def-measure-vector-U}
U = \left\{
\begin{array}{c@{,}c}
A[0] & E[0] \\
A[1] & E[1] \\
\vdots & \vdots \\
A[N_T-1] & E[N_T-1] \\
\end{array}
\right\}  \in \mathbbm{R}^{N_T \times 2} \;,   
\end{equation} 
the matrix of measures. We will refer to the channels $A$ and $E$ as $U_A$ and $U_E$, and to the measures at time $t_n$ as $U_A[n]$ and $U_E[n]$. We will denote by $\alpha$ either channel $A$ or channel $E$. Thus $U_\alpha$ will design the projection of the total dataset $U$ on the channel $\alpha$, \ie the content of the first or the second columns of the matrix of measurements.

For each channel $\alpha \in \{A,E \}$, the measure at time $n$ can be split as:
\begin{equation}
\label{eq:def-signal-noise-decomposition-time}
U_\alpha[n] = U_\alpha^0[n] + N_\alpha[n] \;,
\end{equation}
where $U_\alpha^0[n]$ is the sought-for GW signal emitted by GB sources and received at time $t_n$, and $N_\alpha[n]$ is the noise simultaneously measured.

Consequently the frequency domain is also discretized with a frequency step $\delta f$ depending on the total duration of observation, and we will refer to the discrete frequencies as $f_k = k \delta f$, with $-N_f \leq k \leq N_f$ and $2N_f+1 = N_T$. Since we deal with real-valued signals, their negative ($-N_f \leq k \leq 0$) and positive ($0 \leq k \leq N_f$) frequency Fourier coefficients are complex-conjugated. In the following we will reserve the indices $n \in \{0, \ldots, N_T \}$ and $k \in \{-N_f, \ldots, N_f \}$ for the time and frequency variables respectively.

We will consider an additive noise with the following properties:
\begin{enumerate}
    \item Gaussianity: in time domain and frequency domain, the noise obeys a gaussian law $\mathcal{N}(m,V)$ of mean $m$ and variance $V$.
    \item Stationnarity: the noise \autocorrelation function is left invariant by time translations.
    \item Zero-mean: in Fourier domain, the noise has a null mean value \cite{LDCwebsite}.
\end{enumerate}
Most of what follows can be straightforwardly adapted if we lift the gaussian hypothesis. When needed we will indicate what could be modified to go towards more general (and more realistic) noise descriptions.

For a measurement $U_\alpha$ in channel $\alpha$, we define the  discrete Fourier transform:
$$
\mathcal{F}(U_\alpha[n])[k] =\Delta T \sum_{n = 0}^{N_T -1} U_\alpha[n]    e^{-\frac{2 \i \pi k n}{N_T} } \;,
$$
and the whitened Fourier transform:
$$
\widehat{U}_\alpha [k] = \mathbf{\Sigma}_\alpha^{-1/2}\mathcal{F}(U_\alpha[n])[k] \;,
$$
where $\mathbf{\Sigma}_\alpha$ is the noise power spectral density (PSD), \ie the Fourier transform of the \autocorrelation function. 

We adopt the notation: 
\begin{equation}
\label{eq:def-measure-vector-U-hat}
\widehat{U} = 
\left\{
\begin{array}{c@{,}c}
\widehat{U}_A[-N_f] & \widehat{U}_E[-N_f] \\
\vdots & \vdots \\
\widehat{U}_A[+N_f] & \widehat{U}_E[+N_f] \\
\end{array}
\right\}  \in \mathbbm{C}^{(2N_f+1) \times 2} \;, 
\end{equation}
for the whitened signals in the Fourier domain similarly to our time-domain convention (\ref{eq:def-measure-vector-U}). In the same spirit, the noise whitened Fourier transform for channel $\alpha$ writes:
\begin{equation}
\label{eq:def-noise-fourier-domain}
\widehat{N}_\alpha[k] = \mathbf{\Sigma}_\alpha^{-1/2} \mathcal{F}(N_\alpha[n])[k] \;,
\end{equation}
and the decomposition between whitened physical signal and noise in Fourier domain follows:
\begin{equation}
\label{eq:def-signal-noise-decomposition-frequency}
\widehat{U}_\alpha = \widehat{U}_\alpha^0 + \widehat{N}_\alpha \;.
\end{equation}
Under these hypotheses, the distribution followed by the noise on each channel $\alpha \in \{A,E\}$ in Fourier space explicitly reads:
\begin{equation}
\label{eq:FD_noise_distribution}
\widehat{N}_\alpha[k] \sim \mathcal{N}(0,1) + \i \mathcal{N}(0,1) \;. 
\end{equation}

\section{Sparse signal modelling}
\label{par:FbF_sparsity}

\subsection{Notations}

For a complex-valued series $\{F[l]\}_{l \in \mathbb{Z}}$ and a set of integers $\Lambda \subset \mathbb{Z}$, we classically denote the $L_r(\Lambda)$-norms ($r \geq 1$) as:
\begin{equation}
\label{eq:def-standard-norms}
\norm{F}_{\Lambda; r} = \Big( \sum_{l \in \Lambda} |F[l]|^r \Big)^{1/r} \;,
\end{equation}
where $|F[l]|$ is the modulus of $F[l]$. In the following, it will sometimes be more convenient to write:
\begin{displaymath}
\norm{F[l]}_{\Lambda; r} = |F[l]| \;,
\end{displaymath}
when considering only the term of index $l$ in the series $\{F[l]\}_{l \in \mathbb{Z}}$. The $L_\infty(\Lambda)$-norm is defined by: 
\begin{equation}
\label{eq:def-infty-norm}
\norm{F}_{\Lambda; \infty} = \max_{l \in \Lambda} |F[l]| \;,
\end{equation}
If more generally $F$ takes values admitting a $L_s(\Lambda')$-norm, we introduce the composite norm $L_{r,s}(\Lambda)$:
\begin{equation}
\label{eq:def-composite-norm}
\norm{F}_{\Lambda,\Lambda'; r,s} = \Big(\sum_{l \in \Lambda} \norm{F[l]}^r_{\Lambda'; s}\Big)^{1/r} \;.
\end{equation}
For the sake of conciseness, when $\Lambda$ contains the \emph{support} of the complex-valued series $\{F[l]\}_{l \in \Lambda}$ (\ie the set of indices $l \in \mathbb{Z}$ such that $F[l] \neq 0$), we will simply drop the explicit mention of $\Lambda$ since all indices bringing a \nonzero contribution to the norm will be considered. 

We write $A^*$ the \emph{conjugate transpose} of a matrix $A$ and use $\odot$ for the Hadamard product $C = A \odot B$ of two matrices $A$ and $B$ of the same size:
\begin{equation}
C_{k,l} = A_{k,l} B_{k,l} \;,
\end{equation}
where the indices $k, l$ ranges between 1 and the number of rows or columns of the considered matrices. We keep this notation \textit{mutatis mutandis} in the case of two vectors $A$ and $B$, or one vector $A$ and one matrix $B$, if both have the same number of lines. For example:
\begin{equation}
C_{k} = \{A_{k} B_{k,1}, \ldots, A_{k} B_{k,m} \} \;,
\end{equation}
if $A$ is a vector, $B$ a matrix admitting $m$ columns, and both $A$ and $B$ share the same number of rows.

At last we will denote $\mathbbm{1}_d = \{1,1,...,1\}$ the vector of a $d$-dimensional $\mathbb{R}$ or $\mathbb{C}$-vector space with all components set to 1. When there are no ambiguities, we will drop the index $d$ to keep notations compact.

\subsection{$L_1$ Minimization}
\label{par:L1_minimization}
Detecting and recovering GBs from GW data can be recast as a classical denoising problem. Indeed, the observed data in the channel $\alpha$ are given as a vector of noisy measurements $\mathcal{F}(U_\alpha[n])[k]$. Its content is modeled as a superposition of \quasimonochromatic signals contaminated by noise. Assuming the knowledge of the noise PSD $\mathbf{\Sigma}_\alpha$, the whitened (Fourier transformed) data stream $\widehat{U}_\alpha$ comes as $2N_f+1$ independent and identically distributed realizations of noise $\widehat{N}_\alpha$ on top of the physics content $\widehat{U}_\alpha^0$. We build an estimator of the GB signal $\widehat{U}_\alpha^0$ from the $2N_f+1$ random variables $\widehat{U}_\alpha[k]$ by solving an optimization problem: we search the sparsest representation of the signal $\mathbf{\Sigma}_\alpha^{1/2} \widehat{U}_\alpha^0$ in the Fourier basis $e^{-2 \i \pi k n/N_T}$.

This estimator $\widehat{S}_\alpha$ is actually obtained by minimizing the following cost function:
\begin{eqnarray}
\widehat{S}_\alpha  
& = & \argmin_{\widehat{V} \in \mathbbm{C}^{2N_f+1}} \Bigg[ \norm{\gamma \odot \widehat{V}}_{1} + \frac{1}{2}(\widehat{U}_\alpha-\widehat{V})^* (\widehat{U}_\alpha-\widehat{V}) \Bigg] \nonumber \\
& = & \argmin_{\widehat{V} \in \mathbbm{C}^{2N_f+1}} \Bigg[ \norm{\gamma \odot \widehat{V}}_{1} + \frac{1}{2}\norm{\widehat{U}_\alpha-\widehat{V}}_2^2 \Bigg] \;, \label{eq:L1_formulation}
\end{eqnarray}
where $\{\gamma[k]\}_{-N_f \leq k \leq N_f}$ is a frequency-dependent positive real regularizing parameter. The $L_1$-norm is a standard sparsity-enforcing term and the remaining quadratic term evaluates the discrepancy between the data and the model. This term generally refers to the negative log-likelihood function in statistical inference.

The problem (\ref{eq:L1_formulation}) can be interpreted as a tradeoff between \emph{"finding the sparsest solution possible"}, embodied by the term on the left, and \emph{"finding the solution that fits the measures the best"}, embodied by the term on the right, that evaluates the square distance between the measurement $\widehat{U}_\alpha$ and its Fourier representation $\widehat{V}$. This tradeoff is balanced through the choice of the regularizing parameter $\gamma$.

Since the problem is entirely separable, the estimator $\widehat{S}_\alpha$ can now be computed analytically frequency by frequency. The solution is given by the so-called soft-thresholding operator : 
\begin{equation}
\widehat{S}_\alpha[k] = 
\begin{cases}
\displaystyle \frac{\norm{\widehat{U}_\alpha[k]}_1 - \gamma[k]}{\norm{\widehat{U}_\alpha[k]}_1} \widehat{U}_\alpha[k]
\quad \text{ if } \norm{\widehat{U}_\alpha[k]}_1 > \gamma[k] \\
0 \qquad \text{otherwise}
\label{eq:L1_solution}
\end{cases}
\end{equation}
The regularizing parameter $\gamma$ acts as a threshold since the measure $\widehat{U}_\alpha[k]$ at frequency $k$ will be discarded if its norm is smaller than $\gamma[k]$, and will be replaced by its excess to $\gamma[k]$ otherwise. We will coin the frequency $k$ as \emph{active} if the signal exceeds the threshold $\gamma$, and \emph{inactive} otherwise.
 
 In the following sections, we will lay down a joint analysis of the A and E channels, and further elaborate on the choice of the threshold $\gamma$. However, the structure of the considered problem will always remain the same throughout the article. Only the sparsity-enforcing term and the minimizing algorithm will be sophisticated in order to adapt them to data processing situations of increasing modeling complexity.

\subsection{Combining A and E signals with joint sparsity}
\label{par:L12_minimization}

While the solution stemming from the $L_1$-minimization provides important insights into the algorithmic nature of the problem, it does not benefit from the redundancy of the physical content in the channels A and E, which have been so far treated separately. Indeed the information contained in both channels are highly correlated (the imprint of the same GW is encoded in both channels), whereas the noise is totally uncorrelated. This leads to the following two observations:
\begin{itemize}
    \item A frequency peak of physical origin in one channel will also be present in the other channel, although with different amplitudes. It indeed happens that a peak appears in one channel, but more difficult to detect in the other one.
    \item For a given frequency, the noise realization can be especially high on a channel and low on the other. An independent analysis of the two channels could output a false positive signal on one channel (and only one) for this noise realisation. 
\end{itemize}
 Jointly processing the A and E channels addresses both problems and is the natural extension of the preceding development.
 
 The very structure of the problem of constructing the sparse estimator $\widehat{S}$ is preserved by making use of the compact notation (\ref{eq:def-measure-vector-U-hat}). In Fourier domain, writing $\widehat{U} = \{\widehat{U}_A, \widehat{U}_E \}$ the joint vector of whitened measures, $\widehat{S} = \{\widehat{S}_A, \widehat{S}_E \}$ the sparse signal estimator and $\widehat{V} = \{\widehat{V}_A, \widehat{V}_E \}$ a trial solution, a joint sparse representation of the channels A and E is achieved through:
\begin{eqnarray}
\widehat{S} & = & 
\argmin_{\widehat{V} \in \mathbbm{C}^{(2N_f+1)\text{x}2}} \Bigg[ \norm{\gamma \odot  \widehat{V}}_{1,2} \nonumber \\
& & \quad + \frac{1}{2} \Tr \left\{(\widehat{U}-\widehat{V})^*  (\widehat{U}-\widehat{V})\right\} \Bigg] \nonumber \\
& = & \argmin_{\widehat{V} \in \mathbbm{C}^{(2N_f+1)\text{x}2}} \Bigg[ \norm{\gamma \odot  \widehat{V}}_{1,2} + \frac{1}{2}  \norm{\widehat{U}-\widehat{V}}_{2,2}^2  \Bigg]
\;, \nonumber \\ \label{eq:general_problem_formulation}
\end{eqnarray}
where $\gamma= \{\gamma[k]\}_{-N_f \leq k \leq +N_f}$ still denotes a positive real threshold and the composite norm $\norm{\gamma \odot V}_{1,2}$ actually combines the two channels through:
\begin{eqnarray}
\norm{\gamma \odot \widehat{V}}_{1,2} 
& = & \sum_{-N_f \leq k \leq N_f} \norm{\{\gamma \odot \widehat{V}\}[k]}_2 \nonumber \\
& = & \sum_{-N_f \leq k \leq N_f} \gamma[k] \sqrt{|\widehat{V}_A[k]|^2+|\widehat{V}_E[k]|^2} \;. \nonumber \\
\label{eq:threshold-mixed-norm}
\end{eqnarray}

As in Eq.~(\ref{eq:L1_solution}), the solution is given by:
\begin{equation}
\label{eq:L12_solution}
\widehat{S}_\alpha[k] = 
\begin{cases}
\displaystyle \frac{\norm{\widehat{U}[k]}_{1,2} - \gamma[k]}{\norm{\widehat{U}[k]}_{1,2}} \widehat{U}_\alpha[k] \qquad \text{if } \norm{\widehat{U}[k]}_{1,2} > \gamma[k] \\
0 \qquad \qquad \qquad \text{otherwise}
\end{cases}
\end{equation}
The problem is still separable and can still be solved frequency by frequency, even if we now jointly examines the channels $A$ and $E$. The role played by $\gamma$ is still clearly a threshold, but now it is the \emph{combined output} $\sqrt{|\widehat{U}_A[k]|^2+|\widehat{U}_E[k]|^2}$ of the two channels that should excess the threshold to tag a frequency as \emph{active}. Its source originates from the mixed norm $\norm{\cdot}_{1,2}$  that enters the problem formulation (\ref{eq:general_problem_formulation}), and which is a standard technique in data processing situations where different types of measurements have to be jointly analyzed \cite{VanDenBerg2009}.

\subsection{Fixing the threshold as hypothesis testing}
\label{par:FbF_threshold_choice}

Whether based on the $L_1$ (see Section~\ref{par:L1_minimization}) or the $L_{1,2}$ (see Section~\ref{par:L12_minimization}) sparsity-enforcing terms, building estimators for GB events eventually led to a thresholding operation in the frequency domain. It turns out that fixing the threshold $\gamma$ can be recast as an hypothesis testing problem, which provides important insights into the construction of the estimator itself. This threshold choice can be performed independently for each frequency $k$. We classically define the two hypotheses $H_0$ and $H_1$:
\begin{description}
    \item[$H_0$:] there is no GW signal at frequency $k$;
    \item[$H_1$:] there is a GW signal at frequency $k$.
\end{description}
Since we assumed that the real and imaginary parts of $\widehat{N}_\alpha[k]$ obey independent standard normal distributions, under $H_0$ (only noise) $\norm{\widehat{U}[k]}^2$ admits a chi-square distribution with $2n$ degrees of freedom $\chi^2_{2n}$, and:
\begin{description}
    \item[$n = 1$:] The channels $A$ and $E$ are independently treated, sparsity is enforced using a $L_1$-norm and $\norm{\widehat{U}[k]} = \norm{\widehat{U}_\alpha[k]}_1$ for $\alpha \in \{A, E\}$.
    \item[$n = 2$:] The channels $A$ and $E$ are simultaneously treated, sparsity is enforced using a $L_{1,2}$-norm:, and $\norm{\widehat{U}[k]} = \norm{\widehat{U}[k]}_{1,2} = \sqrt{|\widehat{U}_A[k]|^2+|\widehat{U}_E[k]|^2}$.
\end{description}
Elaborating on Eq.~(\ref{eq:L1_solution}) and Eq.~(\ref{eq:L12_solution}), this provides a criterion to set the threshold value. Adopting \textit{a priori} a probability threshold $p$ --or equivalently a \emph{rejection rate} $\rho = 1-p$-- and defining the real $x_0$ by the value of the cumulative distribution function:
\begin{equation}
\label{eq:x0-from-p-value}
\mathbbm{P}(\chi^2_{2n} \geq x_0) = \rho \;,
\end{equation}
the hypothesis $H_0$ (resp. $H_1$) is adopted for frequency $k$ if $\norm{\widehat{U}[k]}^2 \leq x_0 = \gamma^2[k]$ (resp. $\norm{\widehat{U}[k]}^2 > x_0 = \gamma^2[k]$). 

If the noise distribution is not assumed to be standard normal, then $\norm{\widehat{U}[k]}^2$ will not obey a $\chi^2$ distribution. However, the reasoning above can be adapted to the actual distribution of $\norm{\widehat{U}[k]}^2$, and the principle of fixing the threshold as hypothesis testing remains.

\subsection{Reweighted $L_{1,2}$ minimization}
\label{par:rwgt_L12}

The estimator $\widehat{S}_\alpha[k]$ derived from Eq.~(\ref{eq:L1_solution}) or Eq.~(\ref{eq:L12_solution}) is a rescaling of the noisy signal $\widehat{U}_\alpha[k]$ by a multiplicative factor in $[0, 1]$, which makes it intrinsically biased. We present here an iterative process, named \emph{reweighting} and described in Ref.~\cite{Candes2007}, to correct this bias. It consists in iteratively alternating at each step $m \in \mathbb{N}$ between a resolution phase, yielding  a signal estimator $\widehat{S}^m_\alpha[k]$, and a threshold determination phase, producing a regularizing parameter $\gamma^m[k]$. The convergence of this algorithm has been established in Ref.~\cite{Chen14} and in practice only few (about 3-4) iterations are needed.

As in Section~\ref{par:FbF_threshold_choice}, we set a p-value and define a (square of) threshold $x_0$ through Eq.~(\ref{eq:x0-from-p-value}). We initialize the iterative procedure with the following threshold:
\begin{equation}
\label{eq:def-threshold-first-step}
\gamma^0 = \sqrt{x_0}\mathbbm{1} \;.
\end{equation}
We target a residual $\epsilon_\gamma > 0$ which will act as a stopping criterion: the algorithm will stop at the first step $m$ such that $\norm{\gamma^{m+1}-\gamma^{m}}_\infty < \epsilon_\gamma$. Each iterate of the reweighting procedure performs as follows :
\begin{description}
\item[Phase 1]: Using $\gamma^m[k]$, compute the estimator $\widehat{S}^m$ at step $m$ from Eq.~(\ref{eq:L12_solution}).
\item[Phase 2]: Using $\widehat{S}^n$, evaluate the threshold $\gamma^{m+1}[k]$ from:
\begin{equation}
\label{eq:threshold_L12_rwgt}
\gamma^{m+1}[k] = \frac{(\gamma^0[k])^2}{\kappa \norm{\displaystyle \widehat{S}^m[k]}_{1,2} + \gamma^0[k]} \;,
\end{equation}
where $\kappa$ is a positive real parameter that is set either to amplify ($\kappa > 1$) or to reduce ($\kappa < 1$) the basic reweighting ($\kappa = 1$). Using $\kappa \in \{1, \ldots, 10\}$ yields quantitatively similar results, and we have set $\kappa = 3$ throughout this study.
\end{description}

\begin{figure}[h!]
    \centering
    \includegraphics[width=0.5\textwidth]{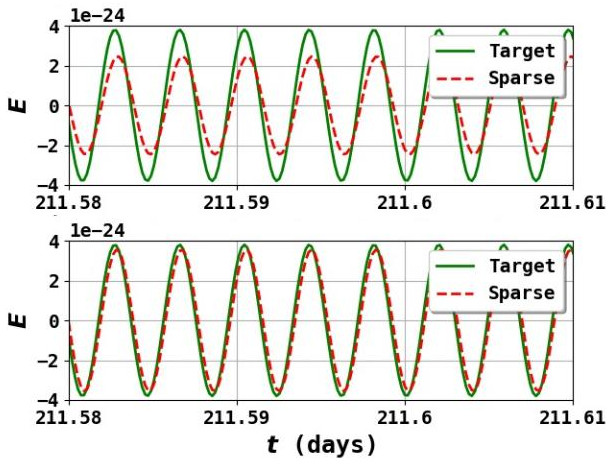}
    \caption{GW signals on channel $E$ as a function of time: target noiseless signal $U_E[n]$ (green) and sparse reconstruction $S_E[n]$ (red). Top: signal estimation without reweighting process. The threshold determination causes an underestimation of the signal. Bottom: signal estimation with reweighting. The bias is mostly corrected.}
    \label{fig:comparison_noRWGT_blockRWGT}
\end{figure}

Note that the reweighting procedure impacts only \emph{active} frequencies (\ie the frequencies $k$ satisfying $\displaystyle \norm{\widehat{S}[k]}_{1,2} > 0$). The higher the norm of the solution, the greater is the correction and the lower becomes the bias, as shown in Figure \ref{fig:comparison_noRWGT_blockRWGT}.

\section{Structured sparsity, a block-based analysis}
\label{par:block_sparsity}

\begin{figure*}[!h]
    \centering
    \begin{minipage}{.49\textwidth}
        \centering
        \includegraphics[width=1\textwidth]{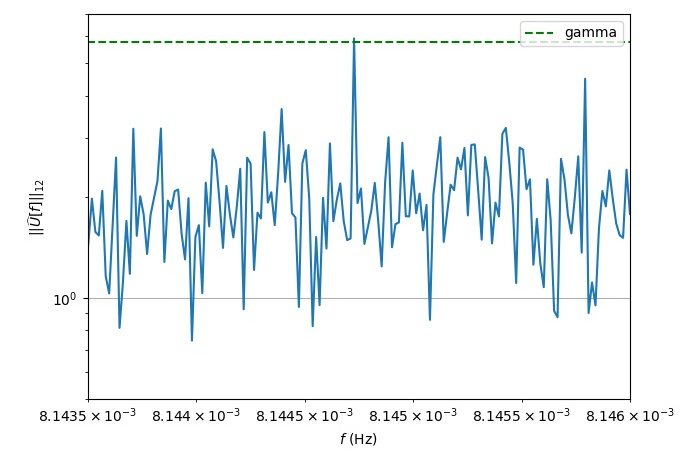}
        \caption{Mixed norm $\norm{\widehat{U}[k]}_{1,2}$ as a function of frequency (solid blue) and compared to the threshold $\gamma$ (dashed green). A noise realisation jumps above threshold and is likely to be detected as a signal, generating a FP frequency.}
        \label{fig:FP}
    \end{minipage}%
    \hfill
    \begin{minipage}{.49\textwidth}
        \centering
        \includegraphics[width=1\textwidth]{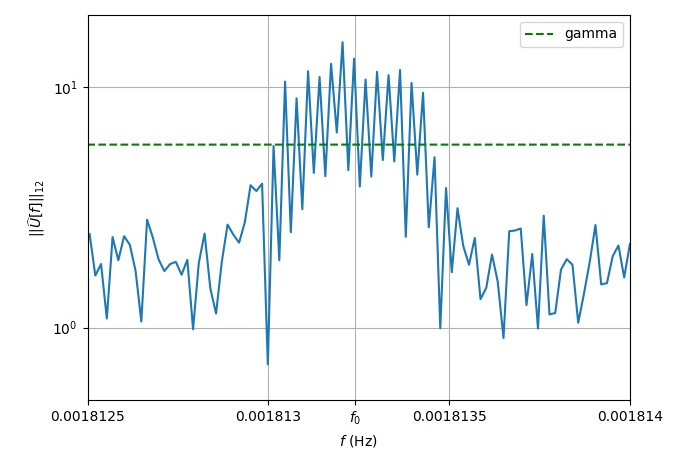}
        \caption{Mixed norm $\norm{\widehat{U}[k]}_{1,2}$ as a function of frequency (solid blue) and compared to the threshold $\gamma$ (dashed green). Within the frequency peak at $f_0$, some harmonics have amplitudes that are lower than the chosen threshold, and thus will not be interpreted as signals, generating FN frequencies.}
        \label{fig:FN}
    \end{minipage}
\end{figure*}

Whether dealing with one or several channels, none of the methods we introduced so far makes use of any physical input on the shape of the GW signal beyond its \quasimonochromatic nature. Since these methods use no information about the structure of GB waveform, they will be further dubbed {\it unstructured} sparsity-based methods. To provide more insight into these methods, we will more specifically consider the following two phenomena:
\begin{itemize}
\item A frequency for which no signal should be detected has been tagged as active. We will refer to it as \emph{false positive} (FP). This situation is illustrated in Fig.~\ref{fig:FP}.
\item A frequency for which a signal should be detected has not been tagged as active. We will refer to it as \emph{false negative} (FN). This situation is illustrated in Fig.~\ref{fig:FN}.
\end{itemize}
In particular, estimating a signal frequency by frequency as in Eq.~(\ref{eq:L1_solution}) or Eq.~(\ref{eq:L12_solution}) leads to a high rate of both FPs and FNs. With respect to the detection of GBs to build a catalogue of sources, the main limitation is the high number of detected FPs. To be able to detect a low amplitude peak, the chosen initial threshold should not be too high. However, the lower the threshold, the higher the probability to get a FP signal. Therefore, an efficient and robust method to separate genuine signals from FPs is much needed. 

This separation can be achieved by implementing the simple yet important remark: a GB signal is not exactly monochromatic, a frequency peak has a characteristic width of a few dozens of frequencies for which the signal has a higher amplitude than the chosen threshold (see Fig.~\ref{fig:FN}). The presence of a signal induces a significant correlation on consecutive frequencies. On the contrary, a FP is often the manifestation of a rare noise realization with a single frequency peak above the threshold (see Fig.~\ref{fig:FP}). Considering the large number of records (the original time series contain tens of millions of data points), the presence of FPs is naturally expected. 

Note that there are also simple mechanisms providing FNs. For example, choosing a frequency-independent threshold leads to the rejection of all frequencies with amplitudes lying too low within the peak (which is likely to happen as amplitude can vary up to nearly 1 order of magnitude within the peak). 

Nevertheless, for FPs as well as FNs, scrutinizing the content of the signal over blocks of neighbouring frequencies appears as a natural solution to solve both problems at once. More precisely, it is unlikely that a noise-related FP activates several contiguous samples. In contrast, a GB signal activates a dozen of Fourier samples. Therefore averaging over the neighborhood of a peak should allow the discrimination of a FP from genuine signal peak.

\subsection{Notations}
\label{par:notations-blocks}

Up to the frequency step $\delta f$ introduced in Section~\ref{par:context}, structuring the frequency domain in blocks amounts to partitioning the set $\{-N_f, \ldots, +N_f\}$ in $I$ successive intervals $B_i$, \ie
\begin{equation}
\label{eq:def-block-partition}
\{-N_f, \ldots, +N_f\} = \bigcup\limits_{1 \leq i \leq I} B_i \; \textrm{with} \; B_i \cap B_j = \emptyset \; \textrm{for} \; i \neq j \;.
\end{equation}
We will refer to an interval $B_i$ as  a \emph{block}. We stress that each $B_i$  is an \emph{ordered set of integers without gaps}: it contains all integers $k$ between the indices referring to the smallest and the largest frequencies considered in the current block. We note $|B_i|$ its cardinal. There exist of course more than one partition of the frequency range $\{-N_f, \ldots, +N_f\}$, and we will identify a partition $\frak{B}$ by its block content:
\begin{equation}
\label{eq:def-partition}
\frak{B} = \{ B_i \}_{1 \leq i \leq I} \;.
\end{equation}
We will similarly note $|\frak{B}| = I$ the number of blocks in the partition $\frak{B}$.

Now recall that a single frequency $k$ has been previously activated if the mixed norm of the data $\norm{\widehat{U}[k]}_{1,2} = \sqrt{|\widehat{U}_A[k]|^2+|\widehat{U}_E[k]|^2}$ exceeds a specific threshold according to Eq.~(\ref{eq:L12_solution}). The rationale behind the so-called {\it structured} sparsity naturally leads to consider:
\begin{equation}
\label{eq:two-two-norm-block}
\sum_{k \in B} \Big( |\widehat{V}_A[k]|^2 + |\widehat{V}_E[k]|^2 \Big) = \sum_{k \in B} \norm{\widehat{V}[k]}_2^2 = \norm{\widehat{V}}_{B; 2,2}^2 \;,
\end{equation}
and to introduce a threshold $\gamma_{\frak{B}}[B]$ for each block $B$ of the partition $\frak{B}$ through:
\begin{equation}
\label{eq:def-mixed-norm-block}
\normBlock{\gamma_{\frak{B}} \odot \widehat{V}}_{\frak{B},1,2} = \sum_{B \in \frak{B}} \gamma_{\frak{B}}[B] \norm{\widehat{V}}_{B; 2,2} \;.
\end{equation}
This is a generalization of Eq.~(\ref{eq:threshold-mixed-norm}), but this time with the mixed norm $\normBlock{\cdot}_{\frak{B},1,2}$ evaluated over the partition $\frak{B}$ instead of the mixed norm $\norm{\cdot}_{1,2}$ evaluated over all frequency samples. 

\subsection{Problem formulation over a block-structured domain}
\label{par:L12_block_sparsity}

Being provided a partition $\frak{B}$ of the measured frequency range $\{-N_f, \ldots, +N_f\}$, building a sparse estimator of the signal requires to solve the optimisation problem:
\begin{equation}
\widehat{S} = \argmin_{\widehat{V} \in \mathbbm{R}^{(2N_f+1)\text{x}2}} \Bigg[\normBlock{\gamma_{\frak{B}} \odot \widehat{V}}_{\frak{B},1,2} + \frac{1}{2}  \norm{\widehat{U}-\widehat{V}}_{2,2}^2  \Bigg] \;.
\label{eq:L12_block_problem_formulation}
\end{equation}
Analogously to problem (\ref{eq:general_problem_formulation}) which was separable frequency by frequency, the problem (\ref{eq:L12_block_problem_formulation}) is separable by block. Moreover, within each block, the problem can be solved frequency by frequency. For a frequency $k$ in block $B$, the solution indeed writes:
\begin{equation}
\label{eq:Block_L12_solution}
\hspace{-1cm} \widehat{S}_\alpha[k] = 
\begin{cases}
\displaystyle \frac{\norm{\widehat{U}}_{B; 2,2} - \displaystyle \gamma_{\frak{B}}[B]}{\norm{\widehat{U}}_{B; 2,2}} \widehat{U}_\alpha[k] \quad \text{if } \norm{\widehat{U}}_{B; 2,2} >  \gamma_{\frak{B}}[B] \\
0 \qquad \qquad \qquad \text{otherwise}
\end{cases}
\end{equation}
which displays the same pattern as the unstructured solution (\ref{eq:L12_solution}). We thus foresee the need for reweighting elaborating on the discussion of Section~\ref{par:rwgt_L12}; this issue will be addressed below in Section~\ref{par:Block_rwgt_L12}. This leaves us with two questions:
\begin{enumerate}
    \item For a given partition $\frak{B}$, what is the best choice for the threshold?
    \item What is the best choice for the partition $\frak{B}$?
\end{enumerate}

To answer the first question, we now consider $\chi^2$-test by block and will assess:
\begin{description}
    \item[$H_0$:] there is no GW signal in block $B$;
    \item[$H_1$:] there is a GW signal in block $B$.
\end{description}
Since we assumed that the real and imaginary parts of $\widehat{N}_\alpha[k]$ obey independent standard normal distributions, under $H_0$ (only noise) $\norm{\widehat{U}[k]}_{B; 2,2}^2$ admits a $\chi^2$-distribution with $4 |B|$ degrees of freedom $\chi^2_{4 |B|}$. Adopting as above a probability threshold $p$ --or equivalently the rejection rate $\rho = 1-p$, and again defining the real $x_0$ by the value of the cumulative distribution function:
\begin{equation}
\label{eq:x0-from-p-value}
\mathbbm{P}(\chi^2_{4 |B|} \geq x_0) = \rho \;,
\end{equation}
the hypothesis $H_0$ (resp. $H_1$) is adopted for block $B$ if $\norm{\widehat{U}[k]}_{B; 2,2} \leq \sqrt{x_0} = \gamma_{\frak{B}}[B]$ (resp. $\norm{\widehat{U}[k]}_{B; 2,2} > \sqrt{x_0} = \gamma_{\frak{B}}[B]$).


\subsection{\BlockTree algorithm}
\label{par:BlockTree}
\label{par:Best_domain_decomposition}

Tree-based block decompositions have been introduced in sparsity-based signal processing methods so as to adapt them better to the structures of the signals to be recovered. In different contexts, tree-based block decompositions have been used in signal and image denoising (see Ref.~\cite{Evers09} and references therein). The approach we will develop below has been inspired by Ref.~\cite{Peyre2009}, but we adapted the advocated top-down building of the dyadic tree to a bottom-up process binding the fate of adjacent frequencies down to their correlation length. 

Following the terminology introduced in Section~\ref{par:L1_minimization}, we will say that a frequency block $B \subset \{-N_f, \ldots, +N_f\}$ is an \emph{active block} (resp. \emph{inactive block}) if, in the sense of the norm $\norm{\cdot}_{B; 2,2}$, the signal is larger (resp. lower) than a given threshold. 

If series of measurements contains only noise, summing over connected blocks will act as an averaging process. If the block is big enough, outliers --which scarcely occur on two frequencies in a row-- will vanish and the FP rate will decrease. On the contrary, if the block is too big, a nearly monochromatic signal may be drowned into the noise and there is the risk that the signal cannot be detected anymore. Since the amplitude of GB signals can be quite close to that of the noise, this case cannot \textit{a priori} be excluded. At last,  if a large block is tagged as active whereas it is not, then its impact of the sparse solution will be proportionally important to its size.  Thus we intuitively understand that active blocks should neither be too small nor too large. 

There is no reasons for all partitions $\frak{B}$ of $\{-N_f, \ldots, +N_f\}$ to yield the same sparse signal estimators. Optimal solutions would stem from an optimization of the overall domain decomposition, which is a NP-hard problem: the computing cost of its resolution with current computers is prohibitively expensive. We hereby propose a suboptimal, yet pragmatic and efficient, solution to the domain decomposition problem.

\begin{figure}[h!]
    \centering
    \includegraphics[width=0.5\textwidth]{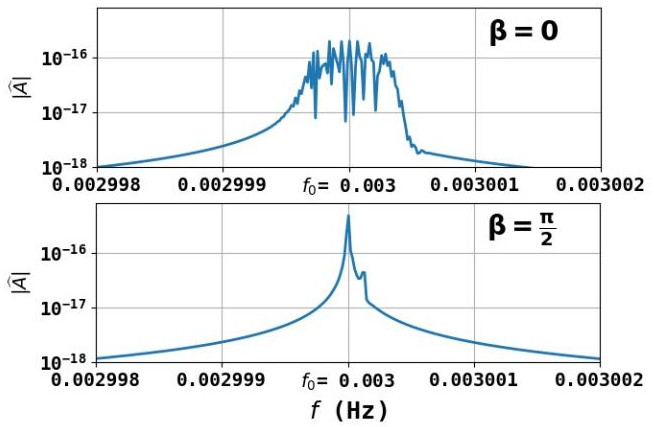}
    \caption{For an observation duration long enough, the peak width is proportional to $\cos(\beta)$, where $\beta$ is the ecliptic latitude for the observed GB. Thus, the largest peak is obtained for $\beta = 0$ (top plot), whereas the thinnest peak is obtained for $\beta = \pi/2$ (bottom plot).}
    \label{fig:waveform_beta0_betaPIsur2}
\end{figure}

To begin with, let us consider a uniform domain decomposition, \ie a partition of $\{-N_f, \ldots, +N_f\}$ made of blocks of the same size. It seems natural to select a block size similar to the width of the peak of the expected GW signal. According to Ref.~\cite{Blaut:2010}, the waveform instant phase is given by:  
\begin{align}
\label{eq:gb-waveform-instant-phase}
\phi(t) = 2 \pi &\bigg(f t +\frac{1}{2} \dot f t^2 \nonumber \\
&+ (f + \dot f t )R \cos(\beta)\cos(\Omega t + \eta_0 - \lambda) \bigg)\;,
\end{align}
where we note $\phi$ the instant phase, $f$ the peak frequency, $\dot f$ the frequency drift, $\beta$ the source latitude in ecliptic coordinates, $\lambda$ the source longitude in ecliptic coordinates, $\eta_0$ the position of LISA around the sun at time $t=0$, $\Omega = 2 \pi / (\text{1 year})$ LISA's orbital frequency and $R$ 1 astronomical unit. Under the assumption $\dot f t << f$, the instant frequency is approximated by:
\begin{equation}
\label{eq:gb-waveform-instant-phase-approximation}
\dot \phi(t) \simeq 2 \pi f- 2\pi f\cos(\beta) R \Omega \sin(\Omega t + \eta_0 - \lambda) \;,
\end{equation}
which means that, for an observation duration long enough ($T_{obs} > 1$ year), the observed half peak width in Fourier domain at first order is given by $|\cos(\beta) R  \omega \Omega|$. Therefore, two extreme cases can be highlighted, which we will focus on during our study:
\begin{description}
    \item[Widest peak] It is obtained by choosing $\beta = 0$. A typical example is displayed on the top row of Fig.~\ref{fig:waveform_beta0_betaPIsur2}. 
    \item[Thinnest peak] It is obtained by choosing $\beta = \pi/2$. A typical example is displayed on the bottom row of Fig.~\ref{fig:waveform_beta0_betaPIsur2}. 
\end{description}
Both peaks are presented with parameters indicated in Appendix~\ref{appendix:GB_parameters}.

Thus a minimal block size $|B| \simeq 10$ is adapted to the characteristic width of the signals we are looking for with a sampling period $\Delta T = 15$s and a total observation period of about 2 years. However, such a small block size do not allow a efficient enough averaging process. This calls for the possibility of merging adjacent blocks to foster this averaging. For that purpose we developed the \BlockTree algorithm, which is a bottom-up approach to frequency domain decomposition:
\begin{description}
\item[Initialisation] Start from a uniform decomposition of blocks with minimal size.
\item[Iteration $m$] Try to merge adjacent blocks 2 by 2 (only if they are of similar sizes):
\begin{itemize}
    \item If a signal was detected in one of the blocks at the iteration $m-1$, but is not detected at the current one: this signal was a FP, so we keep the block resulting of the merging.
    \item If a signal was detected in one of the blocks at the iteration $m-1$, and is still detected at the current one: this signal was not a FP, so we do not merge blocks.
    \item If no signal was detected in both blocks at the iteration $m-1$, we merge blocks.
\end{itemize}
\end{description}
The detailed implementation is described in Algorithm~\ref{algo:BlockTree}. The first iteration aims at cutting drastically the number of FPs. An illustration of the algorithm behavior is shown in Fig.~\ref{fig:blockTree}. The parameters that are usually used for this code are summarized in Appendix~\ref{appendix:algorithm_parameters}.

\begin{algorithm}[h!]
\label{algo:BlockTree}
\SetAlgoLined
\KwResult{\BlockTree }
\KwData{$p$: threshold probability}
\KwData{$n_B$ : minimal block size}
\KwData{$tree_0$: initial tree of identical size blocks (size $n_B$)}
\KwData{$\gamma[B, p]$: threshold, function of block $B$ and probability $p$}
\KwData{$R_{\textrm{comp}}$: Comparability ratio: max ratio until which we consider that 2 blocks have comparable sizes}
 \While{$\big( tree_{n-1} \neq tree_{n} \big)$}{
  
  \eIf{First Iteration}{
      \# Try to group elementary blocks by 4; \\
      For blocks $B_k,B_{k+1}, B_{k+2}, B_{k+3}$: \\
      $B = \cup_{i=k}^{k+3} B_i$ \\
      Compute $D = \norm{\widehat{U}}_{B; 2,2}$; \\
   \eIf{$D < \gamma[B, p]$}{
       \# Try to group elementary blocks by 2; \\
       Merge $B_k,B_{k+1}, B_{k+2}, B_{k+3}$ into $B_k$;
       }{
       $B_1 = B_k \cup B_{k+1}$ ;\\
       $B_2 = B_{k+2} \cup B_{k+3}$;\\
       Compute $D_1 = \norm{\widehat{U}}_{B_1; 2,2}$; \\
       Compute $D_2 = \norm{\widehat{U}}_{B_2; 2,2}$;\\
       \If{$D_1 < \gamma[B_1, p]$}{
           Merge $B_k, B_{k+1}$ into $B_k$;
       }
       \If{$D_2 < \gamma[B_2, p]$}{
           Merge $B_{k+2}, B_{k+3}$ into $B_{k+2}$;
       }
      }
      \textbf{GO TO} next 4 blocks;
   }{
   \#Try to group blocks 2 by 2; \\
   For blocks $B_k$, $B_{k+1}$:\\
   \If{$\displaystyle \frac{\max(|B_k|,|B_{k+1}|)}{\min(|B_k|,|B_{k+1}|)} < R_{\textrm{comp}}$}{
   
       $B = B_k \cup B_{k+1}$ ;\\
       Compute $D = \norm{\widehat{U}}_{B; 2,2}$; \\
       \If{$D < \gamma[B, p]$}{
        Merge $B_k, B_{k+1}$ into $B_k$;\\
       }
       }
       \textbf{GO TO} next 2 blocks
  }
 }
 \caption{\BlockTree}
\end{algorithm}

\begin{figure}
    \centering
    \includegraphics[width=0.5\textwidth]{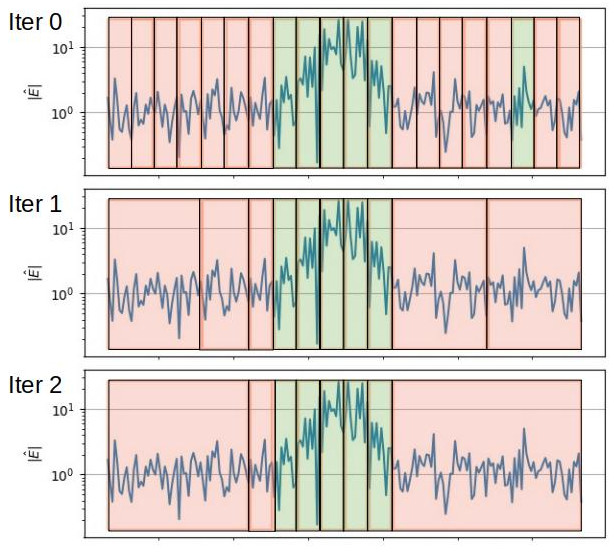}
    \caption{Illustration of the \BlockTree algorithm. We use the notations of Algorithm~\ref{algo:BlockTree}. Evolution of the domain decomposition in a simple case. The red blocks represent inactive blocks, \ie those which norm do not exceed the block threshold. The green blocks represent active blocks, \ie those which norm exceeds the block threshold. The top row represents the initial block decomposition, where all blocks have the same size $n_B$. There are a signal peak in the middle and a FP on the right. This FP will be discarded during the first iteration (middle row). The bottom row represents the final block decomposition. The two blocks on the left side are not merged because their sizes are too different (their ratio is larger than $R_{\textrm{comp}}$).}
    \label{fig:blockTree}
\end{figure}

The \BlockTree approach brings two major improvements:
\begin{enumerate}
    \item We can set a threshold linked to a probability much lower than the one chosen in the case of unstructured sparsity since we are able to discard FPs, paving the way to the detection of more GB signals. 
    \item The easier detection of low amplitude signals enhances the robustness of signal detection.
\end{enumerate}
These two points will be quantitatively assessed in Section~\ref{par:performances_benchmark}.


\subsection{Reweighted $L_{1,2}$ Block Sparsity}
\label{par:Block_rwgt_L12}

We now adapt the discussion of Section~\ref{par:rwgt_L12} to block-structured data. Reweighting will not participate to the tagging of a block as active or inactive but it will just correct the bias introduced\footnote{This is similar to the introduction of a bias through Eq.~(\ref{eq:L1_solution}) and Eq.~(\ref{eq:L12_solution}) as made manifest by the common structure of these thresholding equations.} by Eq.~(\ref{eq:Block_L12_solution}). There are two main options to extend the previous discussions to frequency blocks:
\begin{enumerate}
    \item Set a unique threshold for each block.
    \item Set a frequency-dependent threshold within each block. 
\end{enumerate}
The second option arises from the observation of  amplitude variations between consecutive peaks (see Fig.~\ref{fig:GB_waveform}).

\subsubsection{Block threshold reweighting}
\label{par:block_reweighting}

Similarly to Section~\ref{par:rwgt_L12}, a reweighting procedure can be implemented to correct for the bias induced by the proposed block thresholding. This is described by following steps: 

\begin{description}
\item[Phase 1] Using $\gamma_{\frak{B}}^m[B]$, compute the estimator $\widehat{S}^m$ at step $m$ from Eq.~(\ref{eq:L12_solution}).

\item[Phase 2] Using $\widehat{S}^m$, compute the new threshold $\gamma_{\frak{B}}^{m+1}[B]$ as follows:

\begin{equation}
\label{eq:threshold_L12_rwgt_block}
\gamma_{\frak{B}}^{m+1}[B] = \frac{(\gamma_{\frak{B}}^0[B])^2}{\kappa \norm{\displaystyle \widehat{S}^m}_{B; 2,2} + \gamma_{\frak{B}}^0[B]} \;.
\end{equation}

Following Section~\ref{par:rwgt_L12}, $\kappa$ is a positive real parameter that allows to control the strength of the reweighting procedure. It is again  fixed to $\kappa = 3$ for the block threshold procedure.
\end{description}

The initial block threshold $\gamma_{\frak{B}}^0[B]$ is chosen as a threshold for a $\chi^2_{4 |B|}$ test as explained in Section~\ref{par:L12_block_sparsity}. With the same notations, the choice of threshold is given by $\gamma_{\frak{B}}^0[B] = \sqrt{x_0}$.

\subsubsection{Frequency-by-frequency reweighting for block sparsity}
\label{par:blocktree_unstructured_reweighting}

A potential drawback of the global block reweighting is that the correction factor, which appears in Eq.~(\ref{eq:threshold_L12_rwgt_block}), is identical for each entry of active blocks. Therefore, this might not be as effective as the entrywise reweighting introduced in Section~\ref{par:rwgt_L12} to unbias the thresholding procedure. As shown there, an entrywise reweighting scheme allows to adapt to the amplitude of individual entries of the estimated signal $\widehat{S}$. 
In order to get the best of both the block-based approach and the frequency-based reweighting, we further propose to perform both alternately. In brief, a first estimate $\widehat {S}$ is computed using the block-based thresholding procedure. This allows to carefully account for FPs, and identify active/inactive blocks. The exact same frequency-based reweighting introduced in Eq.~(\ref{eq:threshold_L12_rwgt_block}) is further applied to the samples of the active blocks. This allows to preserve the detection performances of the block-based procedure while significantly enhancing the quality of the detected signals.

Thus, two rejection rates are to be considered: 
\begin{enumerate}
    \item the first one is associated to the \BlockTree, and therefore will be used to sort out real signals and FPs. Typically, we will chose $\rho_{tree} \in \{10^{-5}, 10^{-6} \}$.
    \item The second rejection rate is used for the frequency-based reweighting. Since we already selected the frequencies to keep, we will chose a rejection rate rather high, typically $\rho_{reweighting} = 0.5$ (we would like to keep active most of the frequencies that have been pre-selected with the \BlockTree).
\end{enumerate}

\section{Performances Benchmark}
\label{par:performances_benchmark}

The code that we used for this study is released as an open source code. More information can be found in Appendix \ref{appendix:opensource_code}.

\subsection{Quality checking tools}

Several criteria will be used to assess the quality of the solutions. The detectability of a signal, and its reconstruction, greatly depend on the noise realisation. Thus we will benchmark our algorithms with multiple noise realisations.

\paragraph{Normalized MSE:} The Normalized Mean Squared Error ($NMSE$) gives an estimation of the distance between the real solution $U^0$ (the real signal without noise) and its sparse estimate $S$ in the time domain. The $NMSE$ is defined as follows:

\begin{equation}
\label{eq:def_MSE_global}
    NMSE = -10 \log_{10} \left[\frac{ \norm{U_A^0 - S_A}_2^2 + \norm{U_E^0 - S_E }_2^2}{\norm{ U_A^0 }_2^2+\norm{ U_E^0 }_2^2}\right] \;.
\end{equation}

Note that, by construction, the NMSE is \emph{large} when the sparse estimate $S$ is \emph{close} to the real solution $U^0$. The $NMSE$ gives an idea of the global quality of the reconstructed signal. However, it does not provide any information about the FP and FN rates.

\paragraph{FP rate:} The number and the amplitude of FPs can greatly impact the signal restitution. For a given threshold $p$, we can estimate the FP rate as:

\begin{equation}
    R_{FP}(p) = \frac{\text{\#FP}}{N_f} \;,
    \label{eq:def_FP_rate}
\end{equation}
where \#FP denotes the number of FP frequency peaks.

\paragraph{FN rate:} 

We can then define a FN rate as follow: for a given threshold $p$ for which the FP rate is low, the FN rate will be a function of the amplitude of the input signal (without noise). 
It  will  be  obtained  as an average over $N_{Noise}$ different  noise  realisations. For each input signal amplitude and each noise realisation, one will assess whether the signal was detected by the algorithm. Then, the FN rate will be defined as the proportion of experiments (noise realisations) for which the signal was not detected:
$$
R_{FN}(a) = \frac{1}{ N_{Noise}} \sum_{i=1}^{N_{Noise}} \epsilon_i \;,
$$
where $\epsilon_i = \begin{cases}
1 \text{ if signal with amplitude a } \\
\qquad \text{not detected with }\\
\qquad \text{i-th noise realisation} \\
0 \quad \text{ otherwise}
\end{cases}$

\paragraph{Rejection Rate}: In the remaining of the article, we will refer to the rejection rate $\rho$ defined by:
\begin{equation}
\mathbbm{P}(H_0 \text{ is rejected under } H_0) = 1 - p = \rho 
\label{eq:rejection_rate}
\end{equation}
under the reasoning made in Sections~\ref{par:FbF_threshold_choice} and \ref{par:L12_block_sparsity} where $p$ refers to the threshold probability introduced before and $H_0$ to the absence of detection of a GW signal for a given frequency $k$ or block $B$.

The rejection rate describes the probability to reject $H_0$ for a given frequency or a given block whereas $H_0$ is in fact true. For a uniform domain decomposition (\ie a decomposition frequency by frequency or with uniform blocks), it corresponds to the expected FP rate. Its interpretation is a bit more involved in the case of an adapted domain decomposition, as we will show below.

\subsection{Numerical evaluation with a single galactic binary}
\label{par:perf_gaussianNoise}

\subsubsection{Joint estimation performances}

In order to show the advantage of a joint resolution for channels A and E, we first considered the case of a single GB, which parameters are described in Appendix~\ref{appendix:GB_parameters}, with $\beta = 0$. Two different solutions were considered: the first one is a separate resolution for channels A and E, and the second one is a joint resolution. For both solutions, we made use of the reweighting process. 

\begin{figure}[h!]
    \centering
    \includegraphics[width=0.5\textwidth]{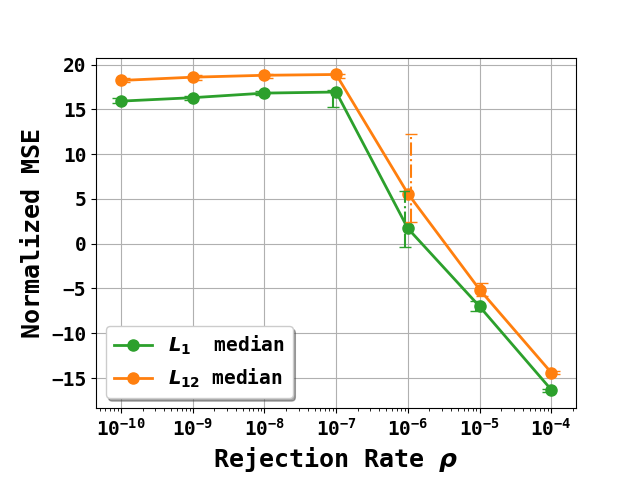}
    \caption{Median $NMSE$ as a function of the chosen probability threshold for a single binary detection with a high amplitude for a single channel resolution ($L_1$-sparsity enforcing, in green) and joint channels resolution ($L_{1,2}$-sparsity enforcing, in orange). The displayed vertical bars correspond to quantiles 0.25 (lower limit) and 0.75 (upper limit) observed for a given number of noise realisations.}
    \label{fig:test0_joint_estimation_performances}
\end{figure}

In Fig.~\ref{fig:test0_joint_estimation_performances}, we plot the median $NMSE$ computed over 25 noise realisations for each displayed point. The green curve represents the solution obtained with a separate resolution, and the orange one is related to the joint resolution. The vertical error bars display the first quartile (quantile of $0.25$) and last quartile (quantile of $0.75$) over the noise realisations. For a high enough rejection rate, the median and the average are quite close and the joint resolution still gives better results than the separate resolution.

We observe that the joint resolution yields a $NMSE$ that is, on average, more than $2.3$ dB better than for the separate resolution.

For the following study, we will only focus on the joint resolution, leaving the separate resolution aside.

\subsubsection{Estimation of the FP rate}

We argued in Section~\ref{par:Best_domain_decomposition} that the uniform domain decomposition (be it frequency by frequency or uniform decomposition) presents a high FP rate compared to the adapted domain decomposition given by the \BlockTree. 

\begin{figure}[h!]
    \centering
    \includegraphics[width=0.5\textwidth]{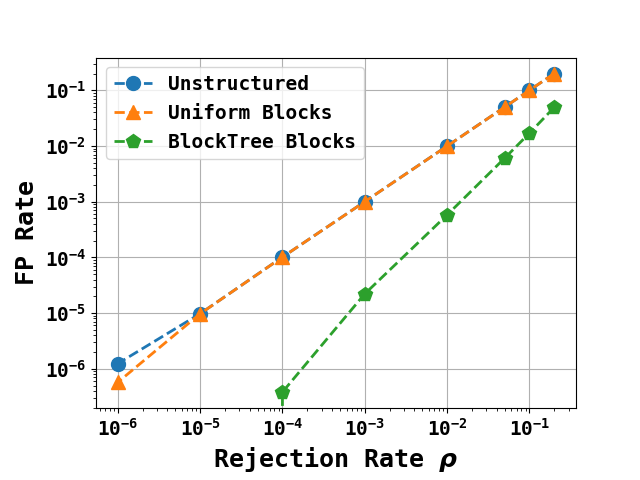}
    \caption{Evolution of the FP rate as a function of different initial thresholds for different resolution classes: unstructured sparsity (blue), uniform blocks resolution with a block size $|B| = 10$ (orange), adapted blocks \BlockTree resolution based on the previous uniform decomposition (green). The probability threshold is chosen so that under $H_0$, $\mathbbm{P}(\chi^2 > x_0) = \rho$. The expected value is given by the first bisector, which is perfectly matched by the two first methods. The \BlockTree enables to have an effective FP rate that is much lower than the expected value due to the averaging process.  }
    \label{fig:test2_benchmark_FPR}
\end{figure}

To illustrate this point, we plot an estimation of the median FP rate as defined in Eq.~(\ref{eq:def_FP_rate}) over 25 noise realisations, for various rejection rates and the three families of methods: unstructured sparsity, uniform block decomposition and \BlockTree decomposition. Fig.~\ref{fig:test2_benchmark_FPR} shows the evolution of the FP rate with the rejection rate $\rho$.

For the unstructured sparsity and uniform block-based approaches, the FP rate coincides with the rejection rate. This is expected since both domain decompositions are uniform. For the \BlockTree domain decomposition, thanks to the averaging process, the effective FP rate is much lower than the rejection rate. Indeed, the probability threshold remains the same (\ie the block threshold $x_0$ is always computed based on the same probability, its variations from one block to another are only due to the change in block size), but the final number of blocks is much lower than for a uniform decomposition, which explains that less blocks result in a FP.

\subsubsection{Evolution of the $NMSE$ as a function of input signal amplitude}

In order to characterize the sensitivity of the proposed methods with respect to the input signal's amplitude, we investigated the evolution of the $NMSE$ as the amplitude of the input signal while the noise amplitude remains constant. Thus, using the same notations as in Section~\ref{eq:def-signal-noise-decomposition-frequency}, the measured signal corresponds to:
\begin{equation}
\label{eq:amplitude_multiplier}
    \widehat{U}_\alpha = A \widehat{U}_\alpha^0 + \widehat{N}_\alpha \;,
\end{equation}
where $A \in [0,1]$ is coined \emph{amplitude multiplier}.

In the following tests, the signal $\widehat{U}_\alpha^0$ is defined by the GB parameters given in Appendix \ref{appendix:GB_parameters} for $\beta = 0$.

\begin{figure*}[h!]
\centering
\begin{tabular}{c c}
\textbf{Frequency Reweighting} & \textbf{Block Reweighting} \\
 \includegraphics[width=0.5\textwidth]{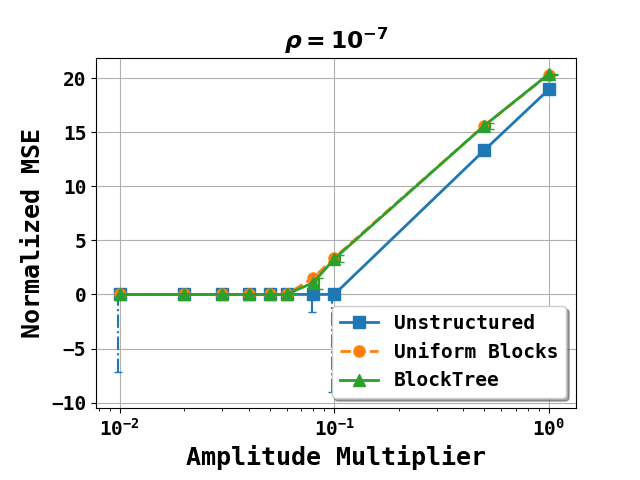}&
\includegraphics[width=0.5\textwidth]{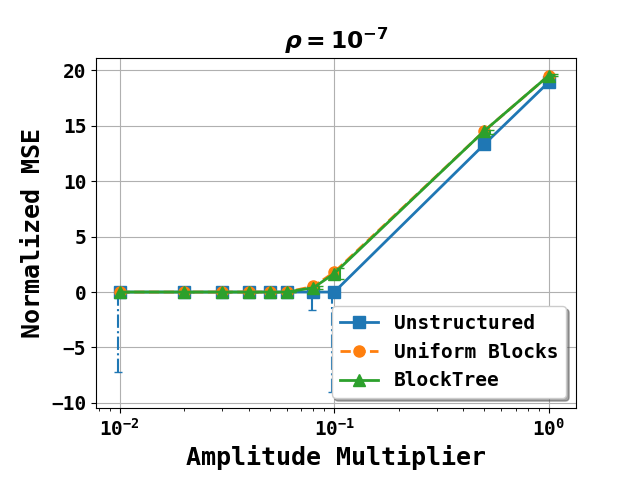}\\
\includegraphics[width=0.5\textwidth]{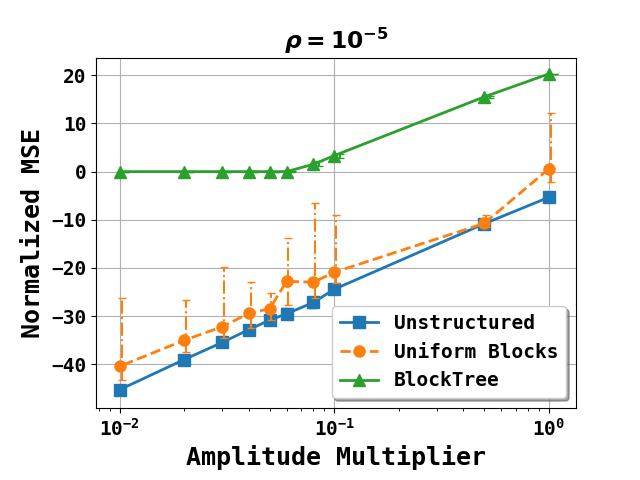}& 
\includegraphics[width=0.5\textwidth]{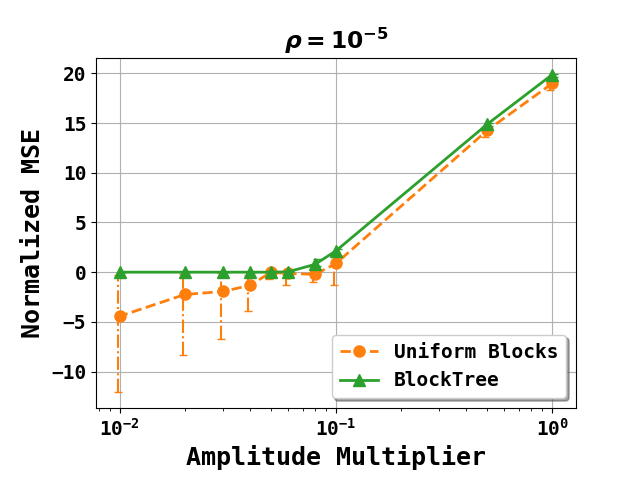}\\
\end{tabular}
\caption{Evolution of $NMSE$ with amplitude multiplier for the 3 main methods presented here: resolution frequency-by-frequency, with uniform domain decomposition and with the \BlockTree domain decomposition. 
Upper row: $\rho = 10^{-7}$ . Lower row : $\rho = 10^{-5}$ . Left: Frequency reweighting as defined in Section~\ref{par:blocktree_unstructured_reweighting}. Right : Block reweighting as defined in Section~\ref{par:block_reweighting}.  The displayed vertical bars correspond to quantiles 0.25 (lower limit) and 0.75 (upper limit) observed for a given number of noise realisations. Due to the important number of FPs for the uniform block decomposition case for the chosen rejection rate $\rho = 10^{-5}$, the $NMSE$ is really low when the signal's amplitude is weak.}
\label{fig:test1_medianMSE}
\end{figure*}

In Fig. \ref{fig:test1_medianMSE}, the results are presented for two different rejection rates: $\rho \in \{10^{-5}, 10^{-7}\}$. For the the lowest rejection rate (upper row), all methods did not present any FPs. In this case, we can see that the block decomposition methods are pretty much equivalent, and provide a better result than the unstructured sparsity method. Moreover, the frequency by frequency reweighting performs better than the block reweighting. 

However, we would not be able to use such a low rejection rate if we want to be able to detect signals with low amplitude, at least for the frequency-by-frequency or uniform block decomposition methods. 

We provide the same plots for a higher rejection rate $\rho = 10^{-5}$ (lower row). Several observations can be done:
\begin{enumerate}
    \item The unstructured sparsity method does the worse performance due to the important number of FPs and, to a lesser extent, the number of FNs.
    \item The \BlockTree domain decomposition may detect signals with lower amplitude than for $\rho = 10^{-7}$, but the limit amplitude multiplier is not that much lower than in the previous case 
    \item The \BlockTree domain decomposition does not present any FPs, as expected from Fig.~\ref{fig:test2_benchmark_FPR}
    \item Using block reweighting limits the impact of FPs: indeed, the number of FPs is identical between frequency and block reweighting, but the $NMSE$ is much better in the case of block reweighting. This is explained by the fact that the bias correction depends on the block norm (close to noise level) in the block reweighting whereas it depends on the frequency amplitude (that can be huge for an outlier) in the frequency reweighting.
\end{enumerate}

\subsubsection{FN rate}
\begin{figure}[h!]
    \centering
    \includegraphics[width=0.5\textwidth]{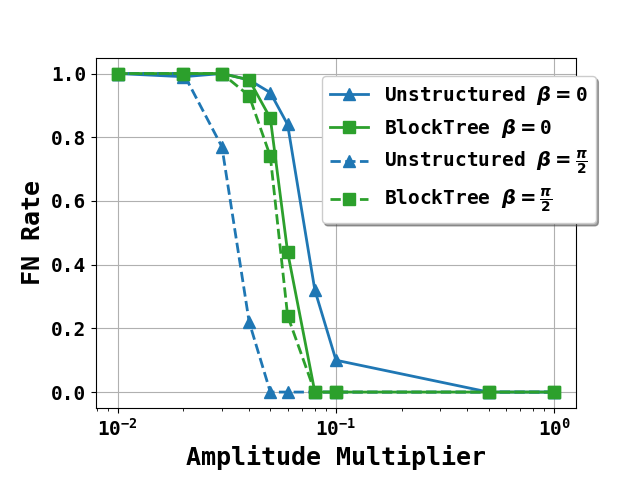}
    \caption{Evolution of FN rate with signal amplitude for extreme peak width. Robustness for our algorithm: the detection rate does not depend on the peak width.}
    \label{fig:test1_FN_rate}
\end{figure}

In Fig.~\ref{fig:test1_FN_rate} is represented the evolution of the FN rate with the amplitude multiplier defined in Eq.~(\ref{eq:amplitude_multiplier}) for a given frequency peak with two different widths: the max peak width ($\beta = 0$) and the min peak width ($\beta = \pi/2$) for the GB parameters described in Appendix \ref{appendix:GB_parameters}.

The frequency-by-frequency performance greatly depends on the peak width: for a given amplitude multiplier, for instance $A = 5.10^{-2}$, for $\beta = \frac{\pi}{2}$ the peak is never detected but for $\beta = 0$ it is always detected. 

This is not the case for the \BlockTree domain decomposition: the detection rate is nearly independent on the peak width, which gives a certain robustness to our algorithm. The uniform block decomposition is an intermediate case between the two methods. 

One could be surprised by the FN rate of the \BlockTree algorithm, that can seem rather low. 
While building the algorithm, we made the choice to favor having the less FPs possible instead of having the less FNs possible, since we would rather have fewer but safer data than more but more uncertain data. 
This explains that our FN rate is not the best among all techniques that we tried, whereas the FP rate shown in Fig~\ref{fig:test2_benchmark_FPR} goes in favor of our algorithm robustness.

\subsubsection{Sensivity to the minimal block size}

As explained in Section \ref{par:Best_domain_decomposition}, the block size can have an impact on the signal recovery realised by the algorithm.

\begin{figure}[h!]
    \centering
    \includegraphics[width=0.5\textwidth]{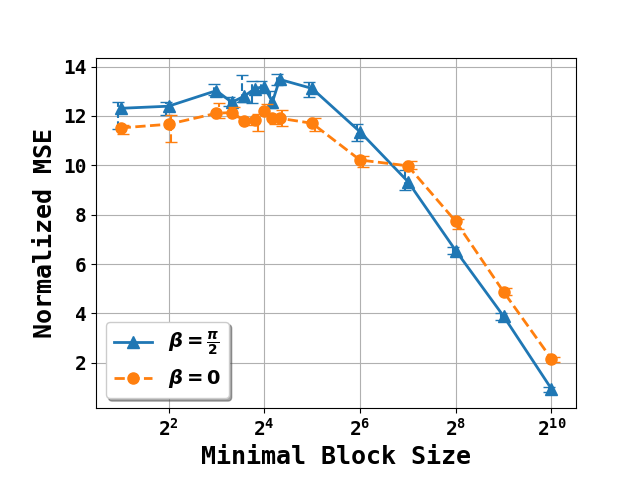}
    \caption{\BlockTree performance for rejection rate $\rho = 10^{-5}$. There is an operating plateau between $|B| = 8$ and $|B| = 20$. The displayed vertical bars correspond to quantiles 0.25 (lower limit) and 0.75 (upper limit) observed for a given number of noise realisations}
    \label{fig:blocksize_sensitivity} 
\end{figure}

Fig.~\ref{fig:blocksize_sensitivity} shows the $NMSE$ sensitivity with respect to the choice of block size $|B|$ for the two extremal peak width cases $\beta = 0$ (widest peak) and $\beta = \pi/2$ (thinest peak). The input signal is given by GB parameters in Appendix~\ref{appendix:GB_parameters} with an amplitude multiplier $A = 0.3$ for a rejection rate $\rho = 10^{-5}$.

There is an operating plateau between $|B| = 8$ and $|B| = 20$ for this specific example. Our choice went for $|B| = 10$.

\subsection{Application to LISA Data Challenges}
The LISA Data Challenges (the data sets and their description) can be found at \cite{LDCwebsite}.

\begin{figure}[h!]
    \centering
    \includegraphics[width=0.5\textwidth]{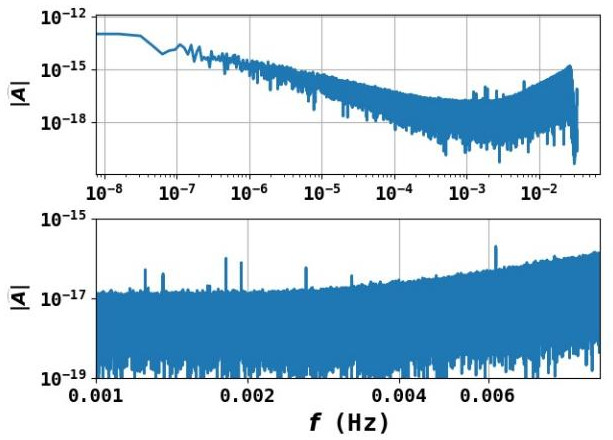}
    \caption{LDC1-3: Verification Galactic white-dwarf binaries : 10 verification binaries have to be found in this signal which noise was generated with the LISA Code noise generator}
    \label{fig:LDC1-3_input}
\end{figure}

\subsubsection{PSD estimator}
For the LISA Data Challenge 1 (LDC1), the noise was created using the LISA Code noise simulator \cite{LDCwebsite}. 
This code simulates a Gaussian noise in time domain, then realises a sub-sampling and a high frequency filtering in the spirit of  what will be realised for LISA real data. The MLDC code provides an estimation of the noise power spectral density (which we will refer to as \emph{theoretical PSD} in the following) that does not take the last two processing steps into account (\ie the sub-sampling and the filtering). 

The proposed methods are highly sensitive to the PSD (through both data whitening and threshold choices). Consequently, the straightforward use of the theoretical PSD did not give satisfying results. Indeed the real and imaginary parts of each signal in Fourier domain whitened by the theoretical PSD $\widehat{A}$ or $\widehat{E}$ is expected to obey a standard normal law $\mathcal{N}(0,1)$. It turns out that at high frequencies the noise distribution of frequency blocks does not follow the expected $\chi^2$ distribution. This motivated the evaluation of a correction to this theoretical PSD. This is consistent with the observed need for an accurate knowledge of the noise PSD in the LIGO-Virgo context \cite{Chatziioannou:2019zvs}. 

\begin{figure}
\centering
\includegraphics[width=0.35\textwidth]{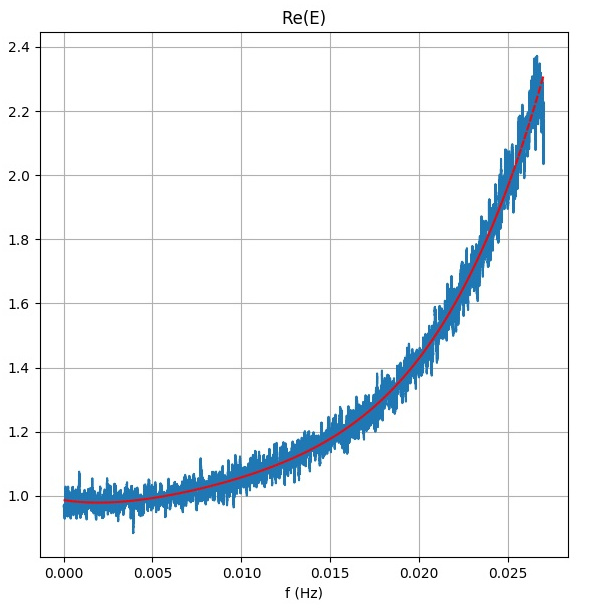}
\caption{In blue: standard deviation $\sigma_{MAD}$ computed with the MAD estimator and a sampling window length of $5000$ points. In red, polynomial fit of order 3 of $\sigma_{MAD}$, referred to as $\sigma_{pol}$.}
\label{fig:LDC1-3_PSD_correction}
\end{figure}

The MAD (Median Absolute Deviation) estimator:
\begin{equation}
\label{eq:def-mad-estimator}
    \sigma_{MAD} = \frac{Median(|X - Median(X)|)}{s_{MAD}} \;,
\end{equation}
with $X$ a window-sample of the initial data and $s_{MAD} \simeq 0.674$, is an empirical estimator of the dispersion of stochastic distributions, which has mainly been advertised in robust statistics \cite{Hampel86}. We use it here to estimate the standard deviation for the real and imaginary parts of both signals $\widehat{A}$ or $\widehat{E}$. The MAD estimate of the standard deviation for $\Re(\widehat{E})$ in presented in Fig.~\ref{fig:LDC1-3_PSD_correction}.

Without the aforementioned sub-sampling and a high frequency filtering effects, this estimator should roughly be equal to 1. As made manifest by Fig.~\ref{fig:LDC1-3_PSD_correction}, this is not the case in practice, and the deviation from the expected value is significant enough to impact the behavior of our algorithm. Thus, for $f < 0.027\textrm{Hz}$ we computed the MAD estimators of the standard deviation of $\Re(\widehat{A})$, $\Im(\widehat{A})$, $\Re(\widehat{E})$ and $\Im(\widehat{E})$ using a sliding sampling window of $5000$ points. We restricted ourselves to $f < 0.027 \textrm{Hz}$ because higher frequencies are too much impacted by the high frequency filtering - this is not a problem since the signals we are looking for have lower frequencies. The size of the sliding window was chosen by finding a trade-off between a window big enough to have an important averaging and a window small enough to have an accurate representation of the correction.

\begin{figure*}
\centering
\begin{tabular}{c c}
\includegraphics[width=0.45\textwidth]{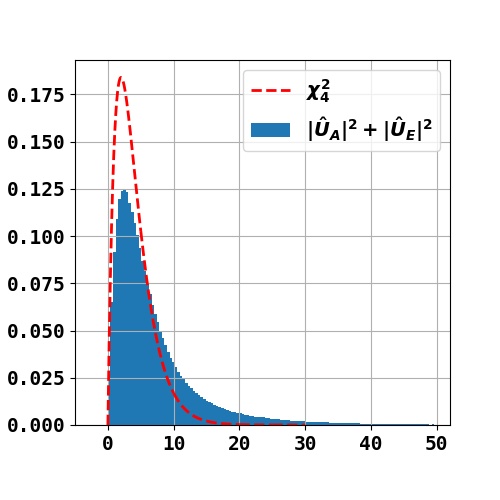}&
\includegraphics[width=0.45\textwidth]{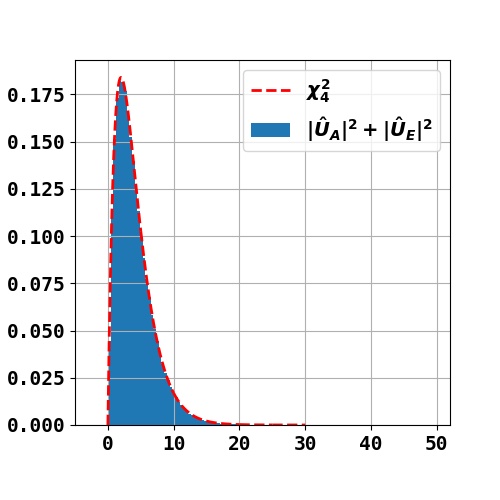}\\
\end{tabular}
\caption{Impact of the PSD correction on the noise distribution. The random variable $\norm{\widehat{U}[k]}_{1,2} = \sqrt{|\widehat{U}_A[k]|^2+|\widehat{U}_E[k]|^2}$ for a frequency $k$ is expected to follow a law of $\chi^2_4$ as explained in Section~\ref{par:FbF_threshold_choice}.
On the left, the signal is normalized with the theoretical PSD $\mathbf{PSD}_{th}$, whereas it is normalized with the effective PSD $\mathbf{PSD}_{eff}$ on the right, for $f < 0.027\textrm{Hz}$. We computed the noise distribution directly on the signal as the amount of signal is negligible compared to the amount of noise. }
\label{fig:LDC1-3_impact_PSD_correction}
\end{figure*}

Each of these four estimators can be fitted by a polynomial of degree 3, and each of these polynomials can be reliably approximated by their average $\sigma_{pol}$. The impact on the whitened noise distribution is illustrated in Fig.~\ref{fig:LDC1-3_impact_PSD_correction}.

To summarize, instead of using the theoretical PSD $\mathbf{PSD}_{th}$, we used the \emph{effective PSD} $\mathbf{PSD}_{eff}$ defined by:
\begin{equation}
    \label{eq:def-effective-psd}
    \mathbf{PSD}_{eff}[f] = \sigma_{pol}^2[f] \mathbf{PSD}_{th}[f] \;,
\end{equation}
to obtain the results presented in the following sections.

\subsubsection{LDC results}

The algorithm parameters used in our code to compute the following results are summarized in Appendix~\ref{appendix:algorithm_parameters}.

\begin{table}[]
\centering
\begin{tabular}{|c|c|c|}
\hline
Peak & $NMSE$ ($\rho = 10^{-5}$) & $NMSE$ ($\rho = 10^{-6}$) \\
\hline 
1 & 9.632 & 9.632 \\
\hline
2 & 9.172 & 9.172 \\
\hline
3 & 4.855 & 4.855 \\
\hline
4 & 2.154  & 2.154 \\
\hline
5 & 15.552 & 15.552 \\
\hline
6 & 13.151 & 13.151 \\
\hline
7 & 14.184 & 14.184 \\
\hline
8 & 5.138 & 5.138 \\
\hline
9 & 3.427 & 2.222 \\
\hline
10 & 13.524 & 13.524 \\
\hline
\textbf{Global} & 12.869 & 12.971\\
\hline
\end{tabular}
\caption{Peak to peak $NMSE$. For $\rho = 10^{-5}$, the algorithm detected 1 FP block (\ie 10 FP frequencies), whereas there was none for $\rho = 10^{-6}$. Even if the local $NMSE$ is better in the case of higher $\rho$, this is not the case for the global $NMSE$ as we detected FP signals.}
\label{tab:peak_to_peak_mse}
\end{table}

\begin{figure*}
\centering
\begin{tabular}{c c}
\includegraphics[width=0.52\textwidth]{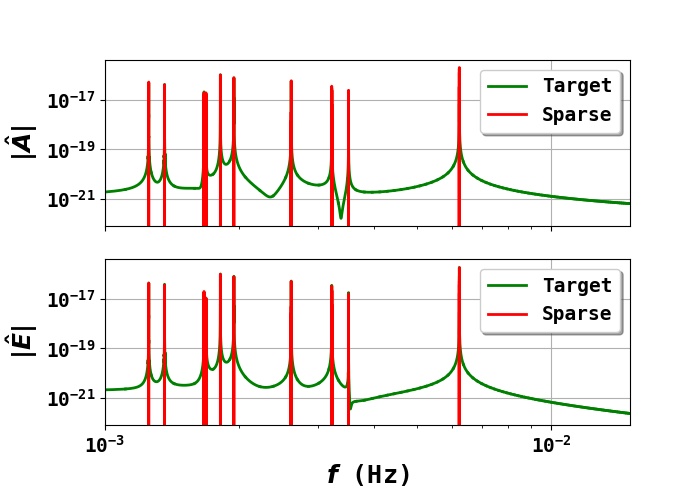}&
\includegraphics[width=0.52\textwidth]{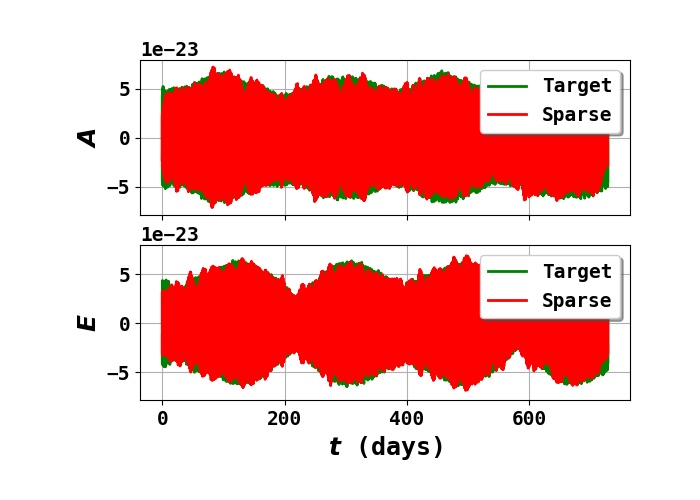}\\
\includegraphics[width=0.52\textwidth]{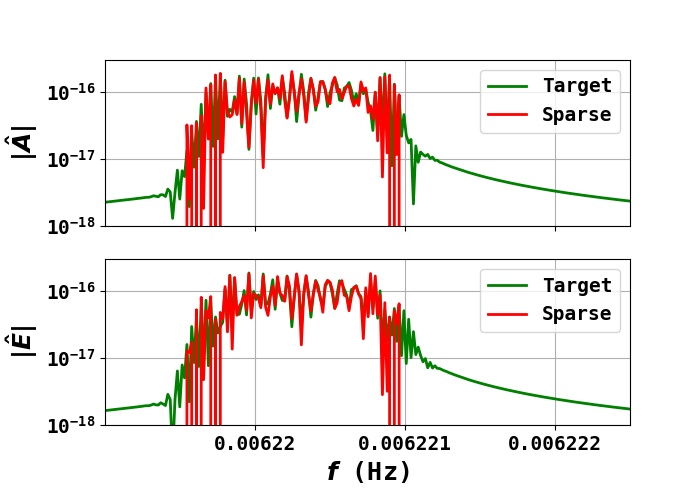}& 
\includegraphics[width=0.52\textwidth]{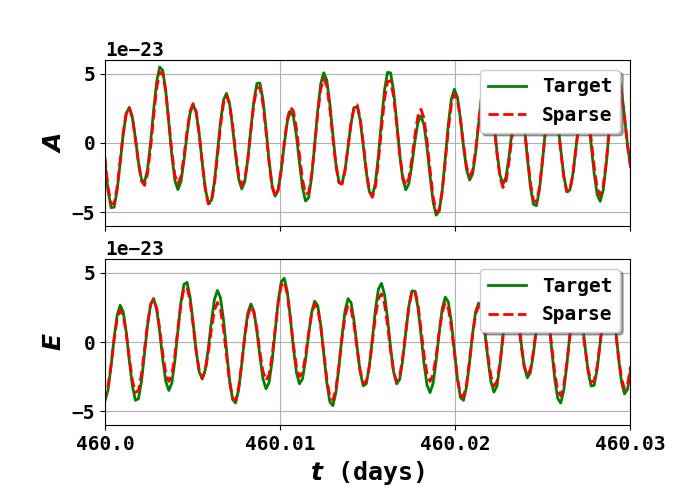}
\end{tabular}
\caption{Results for the LDC challenge with $\rho = 10^{-6}$ and a minimal block size $|B|_{min} = 10$. For this rejection rate, no FP was found. The global solutions are displayed in the upper row, and a zoom was done on the lower row. Left: solutions in Fourier domain, with a zoom on the 10th peak (from the left). Right: solutions in time domain, with a zoom on the solutions. The phases match quite well, the agreement between amplitudes is less satisfactory.}
\label{fig:LDC1-3_results}
\end{figure*}

Fig.~\ref{fig:LDC1-3_results} shows the results that can be expected from the structured sparsity method combined with the \BlockTree algorithm. 
The left column features in upper row the solution found in the Fourier domain in red and the target (real signal without noise) in green: all frequency peaks were recovered. 
The lower plot shows a zoom of the solution on the last frequency peak from the left. This gives an insight on the algorithm behavior: all signals above noise level were recovered. However, this also shows that we have a poor representation of the waveform tails.

The right column upper row represents the corresponding solution in the time domain in red and the target in green. The lower plot is a zoom on the solution on a day portion. 
The phase of the solution matches really well with the one of the target, but the amplitude of the extracted signal is still weaker than the one of the input signal. 
This is to be expected since the basis that we use do not represent the waveform tails in Fourier domain: the energy loss is unavoidable.

\paragraph{Peak-to-peak $NMSE$}

We assessed the performances of the algorithm through estimating a peak-to-peak and a global $NMSE$. 

We define the peak-to-peak $NMSE$ as follow:
\begin{itemize}
    \item We window both the noiseless signal and the extracted signal around the $i^{th}$ peak of interest in frequency domain:  $\widehat{U}_{\alpha,i}^0$,$\widehat{S}_{\alpha,i}$. The window in Fourier domain is defined as follow:
    \begin{enumerate}
        \item We chose a probability threshold ($p_{window}=0.9$ for instance).
        \item We compute the associated threshold for a $\chi^2_4$ law.
        \item We select the regions of the whitened signal that rise above this threshold.
    \end{enumerate}
    Each region corresponds to a frequency peak neighborhood. Thereafter, a frequency peak will be considered as "detected" only if there is at least 1 detected frequency in the detection region defined as before. 
    \item We compute the inverse Fourier transform of the windowed signals: $U_{\alpha,i}^0$,$S_{\alpha,i}$.
    \item We compute the $NMSE$ based on these partial signals:
\end{itemize}

\begin{equation}
\label{eq:def_peak-to-peak_MSE}
    NMSE|_{\text{peak }i} = -10 \log_{10} \left[\frac{ \norm{U_{A,i}^0 - S_{A,i}^0}_2^2 + \norm{U_{E,i}^0 - S_{E,i}^0}_2^2}{\norm{ U_{A,i}^0}_2^2+\norm{ U_{E,i}^0 }_2^2}\right] \;.
\end{equation}

The results are presented in Tab.~\ref{tab:peak_to_peak_mse} for two different rejection rates $\rho \in \{ 10^{-5}, 10^{-7}\}$.

The main difference between the two resolutions is that for rejection rate $\rho = 10^{-5}$, one FP block was detected (\ie 10 frequencies). Choosing a lower rejection rate $\rho = 10^{-7}$ enabled to get rid of it. These results enlight a bit more the trade-off that is done between the FP rate and the resolution quality: even if the peak-to-peak $NMSE$ is better for $\rho=10^{-5}$ for all peaks without any exception, the global performance in this case is not as good as for $\rho = 10^{-6}$, because of the presence of one FP block that greatly affects the measures.

\subsubsection{Residuals study}

\begin{figure}[h!]
\centering
\includegraphics[width=0.5\textwidth]{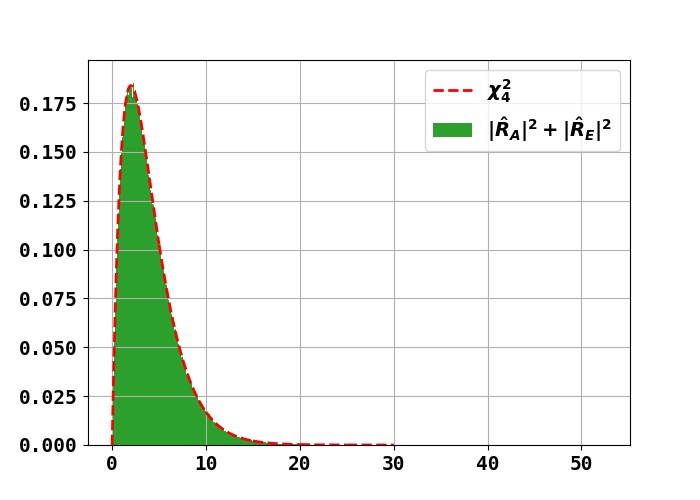}
\caption{In green is represented the distribution of whitened residuals in Fourier domain. In red is plotted the probability density of the law $\chi^2_4$. The residuals manifestly follow the expected law.}
\label{fig:LDC1-3_normed_residuals}
\end{figure}

The residuals for the channel $\alpha \in \{A, E\}$ is defined by :
$$
\widehat{R}_\alpha = \widehat{U}_\alpha - \widehat{S}_\alpha \;,
$$
Fig.~\ref{fig:LDC1-3_normed_residuals} represents the distribution of $\norm{\widehat{R}[k]}_{2,2}^2$ among all frequencies $k$ such that $|k \delta f| < 0.027 \textrm{Hz}$. The residuals seem to follow a law of $\chi^2_4$, which is expected since it is the law followed by the noise (and we only subtracted a few number of frequencies).

\section{Conclusion and prospects}
\label{par:Conclusion}

In this article, we introduced a novel non-parametric framework for the detection and extraction of galactic binary signals from gravitational interferometric measurements. For that purpose, we developed various sparsity-based models and methods that allow accounting for the particular structures of TDI data as well as specificities of galactic binaries' waveform. Namely, the proposed structured sparse model combined with an adaptive tree-based block decomposition, coined \BlockTree algorithm, enables both a control of the false positive rate of the detection procedure as well as a model-independent strategy to capture the main peak of galactic binary waveforms. Combined with a dedicated reweighting procedure, this approach further yields a low-bias estimate of galactic binary signals. 

The proposed approach has been thoroughly evaluated in distinct scenarios involving a single galactic binary event. As well, it has been applied to the detection and retrieval of several galactic binary events from the realistic simulations of the LDC.

The current limitation of the proposed sparsity-based framework is twolfold:
\begin{description}
    \item[Sensitivity to noise modeling] On the one hand, one of the key advantages of such sparsity-based methods is that they come with only few parameters to fix. More precisely, the most important parameter for the detection procedure is the threshold, which can be straightforwardly chosen based on a desired false positive rate. On the other hand, this requires a precise modeling of the noise PSD, which is generally known only approximately. Fortunately, the exact same methodology can be virtually applied to way more general and complex noise models.
    
    \item[Model-independent approach] Let us recall that the proposed framework is model-independent in the sense that no model-based description of the signal's waveform is used in both the detection and estimation procedures. The adaptive block-based decomposition further allows to better retrieve the complex structure of the observed peak of galactic binary signals. However, it is not well-suited to capture the tails of the waveforms, which generally lie below the noise level. It is worth mentioning that the proposed framework can be further extended by substituting the Fourier-based sparse description of galactic binary signals with a dedicated sparse representation of their waveforms.
\end{description}

The proposed sparsity-based framework is quite general, and can serve as a building block to tackle more challenging signal recovery problems from gravitational interferometric measurements. The problem of detecting and recovering galactic binary events has been recast as a sparsity-enforcing denoising problem. We already stressed that it can be extended to account for more general noise models, which can include temporal correlations or non-stationarities. As well, the signal representation is not limited to the Fourier basis. Indeed, sparse signal models can be adapted to a wider range of gravitational events (\eg black hole mergers, etc.) with either fixed basis (\eg Wilson wavelets) or learned dictionaries. The exact same framework can also be adapted to more general inverse problems, including the key problems of missing data or the unmixing of events of different nature.

\section{Acknowledgments}

\label{par:acknowledgments}

 The authors would like to thank S.~Babak, Th.~Foglizzo, A.~Petiteau, and the LISA Data Challenge members for inspiring discussions and warm support. This work was supported by CNES. It is based on the science case of the LISA space mission. JB was supported by the European Community through the grant LENA (ERC StG - contract no. 678282). JB would like to thank the IPARCOS institute of the Universidad Complutense de Madrid for hosting him.

\appendix
\section{Open source code}
\label{appendix:opensource_code}

The code is open source and can be found online at \url{https://github.com/GW-IRFU/gw-irfu} on version 3 of the GPL (GPLv3).

\section{Galactic binary parameters}
\label{appendix:GB_parameters}
The study was conducted choosing the following parameters for the considered GBs:
\begin{description}
\item[Frequency $f_0$] = 3 mHz                       
\item[Frequency derivative $\dot f_0$] = $2.04973995e^{-18} \textrm{Hz}^2$ 
\item[Ecliptic latitude $\beta$] = 0. or 1.570796327 Rad       
\item[Ecliptic longitude $\lambda$] = -2.18009 Rad      
\item[Amplitude $A$] $= 6.44521951e^{-22}$ strain       
\item[Inclination $\iota$] = 0.523599 Rad            
\item[Polarization $\psi$] = 3.61909315 Rad          
\item[Initial phase $\phi_0$] = 2.97459105 Rad          
\end{description}

These parameters are the one needed to create a GB signal by the MLDC code Ref.~\cite{LDCwebsite}.

\section{\BlockTree algorithm parameters}
\label{appendix:algorithm_parameters}
Unless otherwise mentioned, when using the \BlockTree Algorithm~\ref{algo:BlockTree} combined with the unstructured reweighting as in Section~\ref{par:blocktree_unstructured_reweighting}, we use the following parameters:
\begin{description}
\item[$n_B = 10$] (minimal block size)
\item[$R_{comp} = 5$] (comparability ratio)
\item[$\rho_{tree} = 10^{-6}$] (\BlockTree rejection rate)
\item[$\rho_{unstructured} = 0.5$] (reweighting rejection rate)
\item[$\gamma^0 = \mathbbm{1}$] (initial weight in frequency domain)
\item[$\kappa = 3$] (Reweighting coefficient)
\item[$\epsilon_{\gamma} = 0.1$] (convergence criterion for the reweighting algorithm: $\max_k|\gamma_{n+1}[k] - \gamma_n[k]| < \epsilon_\gamma$)
\end{description}

\bibliography{biblio.bib} 
\bibliographystyle{unsrt}

\end{document}